\documentclass[twocolumn]{aastex7}
\usepackage{gensymb}

\usepackage{tabularx}
\usepackage{url}
\usepackage{graphicx}
\usepackage[T1]{fontenc}
\usepackage{ae,aecompl}
\usepackage{booktabs}
\usepackage{natbib}
\usepackage{multirow}
\usepackage{appendix}
\usepackage{overpic}
\usepackage{blindtext}
\usepackage{longtable}
\usepackage{tabu}
\usepackage{lineno}
\usepackage{subcaption} 
\usepackage{chngcntr}
\usepackage{amsmath} 
\usepackage{comment}
\usepackage{xcolor}

\def\aap{A\&A} \def\apjl{ApJ}  
 \def\apjs{ApJS}

\begin{document}

\title{First Detection of Radio Polarization During Jet Formation in the Changing-Look AGN 1ES\,1927+654}

\author[0000-0003-4727-2209]{Onic I. Shuvo}
\affiliation{Department of Physics, University of Maryland Baltimore County, 1000 Hilltop Circle Baltimore, MD 21250, USA}
\affiliation{Astrophysics Science Division, NASA Goddard Space Flight Center, Greenbelt, MD 20771, USA}
\email{oishuvo@umbc.edu}

\author[0000-0002-7676-9962]{Eileen T. Meyer}
\affiliation{Department of Physics, University of Maryland Baltimore County, 1000 Hilltop Circle Baltimore, MD 21250, USA}
\email{meyer@umbc.edu}

\author[0000-0003-2714-0487]{Sibasish Laha} 
\affiliation{Astrophysics Science Division, NASA Goddard Space Flight Center, Greenbelt, MD 20771, USA}
\affiliation{Center for Space Science and Technology, University of Maryland Baltimore County, 1000 Hilltop Circle, Baltimore, MD 21250, USA}
\affiliation{Center for Research and Exploration in Space Science and Technology, NASA/GSFC, Greenbelt, Maryland 20771, USA}
\email{sib.laha@gmail.com}

\author[0000-0003-1101-8436]{Agniva Roychowdhury}
\affiliation{National Centre for Radio Astrophysics, Ganeshkhind, Pune 411007, MH, India.}
\email{agniva@ncra.tifr.res.in}

\author[0000-0002-9163-8653]{Dev R. Sadaula} 
\affiliation{Astrophysics Science Division, NASA Goddard Space Flight Center, Greenbelt, MD 20771, USA}
\affiliation{Center for Space Science and Technology, University of Maryland Baltimore County, 1000 Hilltop Circle, Baltimore, MD 21250, USA}
\affiliation{Center for Research and Exploration in Space Science and Technology, NASA/GSFC, Greenbelt, Maryland 20771, USA}
\email{dsadaula@umbc.edu}

\author[0000-0001-6812-7938]{Tracy E. Clarke}
\affiliation{U. S.\ Naval Research Laboratory, 4555 Overlook Avenue SW, 
Washington, DC 20375, USA}
\email{tracy.e.clarke2.civ@us.navy.mil}

\author[0000-0002-5187-7107]{Wendy M. Peters}
\affiliation{U. S.\ Naval Research Laboratory, 4555 Overlook Avenue SW, 
Washington, DC 20375, USA}
\email{wendy.m.peters8.civ@us.navy.mil}

\author[0000-0003-0936-8488]{Mitchell C. Begelman}
\affiliation{Center for Integrated Plasma Studies, Department of Physics, 390 UCB, University of Colorado, Boulder, CO 80309-0390, USA}
\affiliation{Department of Astrophysical and Planetary Sciences, 391 UCB, Boulder, CO 80309-0391, USA}
\email{mitch@jila.colorado.edu}

\author[0000-0002-2040-8666]{Markos Georganopoulos}
\affiliation{Department of Physics, University of Maryland Baltimore County, 1000 Hilltop Circle Baltimore, MD 21250, USA}
\email{georgano@umbc.edu}

\author[0000-0001-9475-5292]{Rostom Mbarek}
\email{rmbarek@princeton.edu}
\affiliation{Department of Astrophysical Sciences, Princeton University, Princeton, NJ 08544, USA}

\author[0000-0001-9725-5509]{Amelia M. Hankla}
\affiliation{
Joint Space-Science Institute, University of Maryland, College Park, MD 20742, USA}
\affiliation{Department of Astronomy, University of Maryland, College Park, MD, USA}
\email{ahankla@umd.edu}

\author[0000-0001-7801-0362]{Alexander Philippov}
\affiliation{
Department of Physics, University of Maryland, College Park, MD 20742, USA}
\email{sashaph@umd.edu}

\author[0000-0002-5519-9550]{Navin Sridhar}
\affiliation{
Department of Physics, Stanford University, 382 Via Pueblo Mall, Stanford, CA 94305, USA}
\affiliation{Kavli Institute for Particle Astrophysics \& Cosmology, 452 Lomita Mall, Stanford, CA 94305, USA}
\email{nsridhar@stanford.edu}


\begin{abstract}

We present a multiwavelength radio study of the changing-look AGN 1ES\,1927+654 following its renewed X-ray brightening since mid-2022, combining VLBA imaging with multi-band and time-resolved VLA observations. Our main results are: (i) VLBA observations continue to reveal the bipolar jet structure first reported in \citet{meyer2025}, with new X-band detections confirming the morphology at larger separation and K-band imaging revealing a bridge of emission consistent with a continuous outflow; (ii) the VLBA core exhibits a GHz-peaked spectrum, and VLA/VLITE observations from 340~MHz to 45~GHz show emission from an inhomogeneous synchrotron source, self-absorbed below $\sim$3~GHz and optically thin above; (iii) VLA and VLBA flux densities are consistent above 8~GHz, indicating emission dominated by a compact core; (iv) linear polarization is detected above $\sim$3~GHz, with fractional polarization increasing with frequency from $\sim$0.2\% to $\sim$6\% and EVPAs exhibiting non-linear $\lambda^2$ behavior; (v) long-term VLA X-band monitoring reveals a large, smooth rotation of the EVPA of $\sim137^\circ$ over $\sim81$~days in 2026, followed by a partial return, at stable total intensity and suppressed fractional polarization $\lesssim1.4\%$, consistent with a propagating disturbance crossing the compact emitting region; and (vi) circular polarization is not detected at any epoch. These observations demonstrate that 1ES\,1927+654 has transitioned from an X-ray--dominated, non-jetted state to a radio-loud AGN~\citep{laha2025} hosting a newly launched relativistic jet, a unique laboratory for studying jet birth and early magnetic field evolution.

\end{abstract}



\section{Introduction} \label{sec:intro}

Recent advances in time-domain astronomy have revealed a growing population of active galactic nuclei (AGNs) that undergo dramatic and rapid changes in luminosity and spectral appearance. These so-called changing-look AGNs (CL-AGNs) exhibit transitions between optical spectral types on timescales of months to years and often show strong, correlated variability across multiple wavebands \citep{Mathur2018, trakhtenbrot2019, kokubo2020, komossa2020, Komossa2026}. Such behavior challenges the classical AGN unification paradigm, in which differences between AGN types are attributed primarily to orientation and long-lived circumnuclear structures \citep{Antonucci1985, antonucci1993, Bianchi2012, Almeida_ricci2017}. Instead, the rapid transformations observed in CL-AGNs point to intrinsically time-dependent processes, such as sudden changes in accretion rate, disk instabilities, or the formation and destruction of compact coronal structures \citep{Ricci2023}.

Among known CL-AGNs, the formerly type-II source 1ES\,1927+654 stands out as one of the most extreme and well-studied examples \citep{trakhtenbrot2019, ricci21, masterson22, li2022, laha2022, Ghosh2023}. In December 2017, the source underwent a dramatic optical/UV outburst initially compared to a tidal disruption event (TDE), accompanied by the transient appearance of broad optical emission lines. This outburst was followed by the complete disappearance of the X-ray corona between August and October 2018. By late 2018, the broad-line region (BLR) emission had vanished, and the optical/UV light curve entered a decline with $t^{-0.91}$ decay, notably shallower than the canonical $t^{-5/3}$ TDE fallback rate \citep{vanVelzen2021}, leading \citet{laha2022} to propose a magnetic flux inversion event as an alternative origin for the outburst. In November 2019, the X-ray corona reappeared with a luminosity nearly an order of magnitude higher than its pre-flare level, while the UV emission continued to fade. The source returned to a quiescent, pre-outburst state by mid-2021.

Since mid-2022, 1ES\,1927+654 has entered a new phase of activity. The soft X-ray flux has shown a gradual but sustained rise, accompanied by the emergence of a quasi-periodic oscillation (QPO) predominantly in the 2--10~keV band \citep{Masterson2025}. In contrast, the UV emission has remained relatively stable, varying by less than $\sim$30\% \citep{laha2025}. This decoupling between the X-ray and UV bands suggests renewed changes in the innermost accretion flow or coronal structure rather than a simple re-brightening of the accretion disk. More recently, a broad Fe~K emission feature and soft X-ray emission lines have emerged in the X-ray spectrum, absent in all earlier epochs \citep{Sadaula_2026arXiv260705246S}.

Concurrently with this renewed high-energy activity, the source has undergone a remarkable transformation at radio wavelengths. The unresolved radio core flux began an exponential rise in early 2023, increasing by factors of $\sim$40$\times$ and $\sim$60$\times$ at 5 and 8.4~GHz respectively compared to pre-flare levels, with the peak flux of $\sim$70~mJy at 5~GHz reached by mid-2023 \citep{meyer2025}. This rapid brightening represents a rare case of a previously radio-quiet AGN becoming radio-loud on a timescale of only a few months. The spectral index flattened ($|\alpha| < 0.5$ for $S_\nu \propto \nu^{\alpha}$, measured between C and X bands) during the rapid rise and remained flat through most of the following year, while continued VLBA and EVN monitoring revealed a sustained plateau in radio flux. The close temporal agreement between the VLBA/EVN radio peak and the independent monitoring with the Arcminute Microkelvin Imager (AMI) confirms that the radio brightening is intrinsic to the source and not an artifact of calibration, resolution effects, or sparse interferometric sampling.

High-resolution VLBA imaging has provided additional insight into the evolving radio structure of the source. Early VLBA observations at C band in 2021 revealed weak, low-level extended emission on parsec scales, tentatively interpreted as a disk-like or outflow component with a total flux of a few mJy \citep{laha2022}. Follow-up observations in 2022 showed hints of structural evolution, including a possible reduction in the extent of this emission \citep{Ghosh2023}. By April 2023, this extended emission was no longer detected, suggesting it may have originated from an earlier episode of activity that subsequently became optically thin. Indeed, more recent high-frequency VLBA observations have revealed a new compact structure: residual images from K-band data obtained in early 2024 show two components separated by $\sim$0.45~mas ($\sim$0.16~pc) and aligned at a position angle of $\sim$30$^\circ$, implying a mildly relativistic advance speed of order 0.3$c$ \citep{meyer2025}. The high brightness temperatures inferred from the radio emission ($10^{7}$--$10^{9}$~K at 5~GHz) further support an origin in synchrotron emission from this newly launched jet.

The radio spectral energy distribution (SED) of 1ES\,1927+654 has evolved in parallel with the emerging parsec-scale jet structure. During the rapid rise in early 2023, the spectrum was relatively flat between C and X bands, consistent with optically thick synchrotron emission from a compact source. By mid-2023, the spectrum developed a pronounced turnover at gigahertz frequencies, resembling that of gigahertz-peaked spectrum (GPS) sources, which are commonly associated with young or recently triggered radio jets \citep{meyer2025}. Observations with the Submillimeter Array (SMA) at 225 and 345~GHz reveal a flat sub-mm spectrum that is significantly harder than the steep GHz-band jet spectrum at the same epoch, suggesting a physically distinct emission component possibly associated with a compact X-ray--emitting corona.

Building on these prior results, this paper presents new multi-frequency VLBA observations tracing the continued emergence of a parsec-scale jet in 1ES\,1927+654, together with simultaneous VLA multi-band observations and long-term X-band polarization monitoring. Together these data allow us, for the first time, to characterize the radio spectrum, compact jet structure, and fractional polarization properties during the formation of a newborn relativistic jet in a changing-look AGN. This paper is organized as follows: Section~\ref{sec:obs} describes the VLBA and VLA observations and data reduction; Section~\ref{sec:results} presents the imaging, spectral, and polarization results together with a discussion of their implications for jet launching and magnetic field evolution; and Section~\ref{sec:conclusions} summarizes our findings.

\startlongtable
\begin{deluxetable*}{llcrcccccc}
\small
\label{tab:Observation}
\tablecaption{Follow-up VLBA Observations and Radio Properties.}

\tablehead{\colhead{Date} &  \colhead{Segment} &  \colhead{Band} & \colhead{Freq.} & \colhead{RMS}   & \colhead{$F_\mathrm{peak}$} & \colhead{$S_\mathrm{tot}$} & \colhead{Rest.\ Beam} & \colhead{Beam Angle} & Self-calibration\\[-1ex]
        \colhead{}       & \colhead{}     &  \colhead{}	   & \colhead{(GHz)} & \colhead{(mJy bm$^{-1}$)} & \colhead{(mJy\,bm$^{-1}$)} & \colhead{(mJy)}    &        \colhead{($\alpha \times \delta$; mas)}  &  \colhead{(deg)} & \\[-1ex]
        }
        \colnumbers
\startdata
2024-08-12  &BM556D      & L  & 1.61 &0.28  &29.82&39.44  &10.2$\times$5.1 & \phn~$-$33.14&a$+$p\\[-0.8ex]
            &            & C  & 4.87 &0.29  &56.79&56.30 &3.1$\times$1.5 & \phn~$-$24.49&a$+$p\\[-0.8ex]	
            &            & X  & 8.37 &0.12	&56.30 & 55.76  & 1.8$\times$0.9 & \phn~$-$27.59&a$+$p\\[-0.8ex]
            &            & K  & 23.57 &0.14  &5.62 & 6.89  &0.6$\times$0.3 & \phn~$-$27.55& no\\
2024-09-13  &BM571A1     & C  & 4.87 &0.48  &64.15 &67.25 &5.7$\times$1.3 &       \phn~~~1.79&a$+$p\\[-0.8ex]	
            &            & X  & 8.37 &0.34	&50.25 & 54.45  & 3.4$\times$0.7 & \phn~$-$0.72&a$+$p\\
2024-11-08  &BM571B1       & L  & 1.61 &1.42  &45.00&76.1  &9.8$\times$4.8 & \phn~$-$23.88&a$+$p\\[-0.8ex]
            &            & C  & 4.87 &0.52  &66.96&67.86 &3.6$\times$1.9 & \phn~$-$26.46&a$+$p\\[-0.8ex]	
            &            & X  & 8.37 &0.12	&49.09 & 54.43  & 2.1$\times$0.9 & \phn~$-$27.03&a$+$p\\[-0.8ex]
            &            & K  & 23.57 &0.05  &11.07 & 16.25  &0.8$\times$0.4 & \phn~$-$28.28& a$+$p\\
2024-11-29  &BM571A2     & C  & 4.87 &0.14  &81.36 &82.37 &5.4$\times$1.6 &       \phn~~23.49&a$+$p\\[-0.8ex]	
            &            & X  & 8.37 &0.16	&58.37 & 65.57  & 3.8$\times$0.8 & \phn~~20.41&a$+$p\\
2024-12-24  &BM571B2       & L  & 1.61 &0.37  &47.97 &59.32  &10.4$\times$5.3 & \phn~$-$20.76&a$+$p\\[-0.8ex]
            &            & C  & 4.87 &0.15  &77.48 &79.35 &3.7$\times$1.8 & \phn~$-$15.35&a$+$p\\[-0.8ex]	
            &            & X  & 8.37 &0.11	&53.10 & 58.44  & 2.1$\times$0.9 & \phn~$-$18.59&a$+$p\\[-0.8ex]
            &            & K  & 23.57 &0.05  &11.47 & 20.70  &0.8$\times$0.3 & \phn~$-$20.45& p\\
2025-01-31  &BM571A3     & C  & 4.87 &0.27  &92.15 &97.02 &4.8$\times$2.2 &  \phn~~~6.92&a$+$p\\[-0.8ex]	
            &            & X  & 8.37 &0.14	&64.72 &70.85   & 2.9$\times$1.1 &  \phn~~~7.61&a$+$p\\
2025-03-29  &BM571B3       & L  & 1.61 &0.35  &47.26 &63.6  &9.4$\times$3.9 & \phn~$-$19.85&a$+$p\\[-0.8ex]
            &            & C  & 4.87 &0.17  &76.06 &78.43 &3.7$\times$1.8 & \phn~$-$16.85&a$+$p\\[-0.8ex]	
            &            & X  & 8.37 &0.11	&56.29 &62.24   & 2.1$\times$0.9 & \phn~$-$17.46&a$+$p\\[-0.8ex]
            &            & K  & 23.57 &0.12  &13.09 &20.06   &0.8$\times$0.4 & \phn~$-$13.02& p\\
2025-05-22  &BM580       & L  & 1.61 &1.01  &51.30 &49.20  &9.7$\times$4.8 & \phn~$-$25.85&p\\[-0.8ex]
            &            & C  & 4.87 &0.26  &77.40 &81.19 &3.9$\times$1.7 & \phn~$-$29.26&a$+$p\\[-0.8ex]	
            &            & X  & 8.37  &0.12	 &54.98 &60.48   & 2.3$\times$1.2 & \phn~$-$27.91&a$+$p\\[-0.8ex]
            &            & K  & 23.57 &0.18  &15.17 & 22.50  &0.8$\times$0.4 & \phn~$-$27.11& p\\
2025-10-24  &BM588A      & L  & 1.61 &0.84  &64.22 &74.60  &11.2$\times$5.7 & \phn~$-$32.08&a+p\\[-0.8ex]
            &            & C  & 4.87  &0.13  &68.59 &70.45 &3.5$\times$1.7 & \phn~$-$19.82&a$+$p\\[-0.8ex]
            &            & X  & 8.37  &0.08  &41.89 &47.89  &2.1$\times$1.2 & \phn~$-$25.71&a$+$p\\[-0.8ex]
            &            & K  & 23.57 &0.04  &4.90 &15.71   &0.9$\times$0.6 & \phn~$-$33.73&a$+$p\\
\enddata
\end{deluxetable*}

\section{Observations and Data Reduction} \label{sec:obs}

Our analysis is based on a coordinated set of VLBA and VLA observations obtained as part of multiple approved monitoring programs targeting the evolving radio emission from 1ES\,1927+654. The VLBA data were acquired through a sequence of programs spanning semesters 24A through 25B (Legacy IDs: BM556, BM571, BM580, BM588), providing regular snapshots of the source at C, X, and K bands, with additional L-band coverage in later epochs. These observations were designed to track the rapid rise in compact radio flux density, measure the evolving radio spectrum, and resolve newly emerging parsec-scale structures associated with a nascent jet or outflow.

Complementary VLA observations were obtained through dedicated multi-band and monitoring programs (Legacy IDs: AM1812, AM1827, AM1851, AM1883). A simultaneous 1--45~GHz VLA observation in 24A, together with commensal 340~MHz data from the VLA Low-band Ionosphere and Transient Experiment \citep[VLITE;][]{Clarke2016}, provided a complete radio spectrum spanning the spectral turnover, while subsequent X-band monitoring campaigns in semesters 25A, 25B and 26A sampled the evolution of the flux density, fractional polarization, and electric vector position angle (EVPA) on weekly to monthly timescales. Together, the VLBA and VLA datasets offer a uniquely comprehensive view of the spectral, structural, and polarization evolution of 1ES\,1927+654 during the formation of a newborn relativistic jet. We also make use of previously published measurements: VLBA, AMI, and e-MERLIN flux densities prior to mid-2024 from \citet{meyer2025}, and the \textit{Swift}/XRT soft X-ray light curve from \citet{laha2025} and
\citet{Sadaula_2026arXiv260705246S}. No new reduction of these data was performed for this work.

\begin{figure*}[t]
    \centering
    \includegraphics[width=0.95\textwidth]{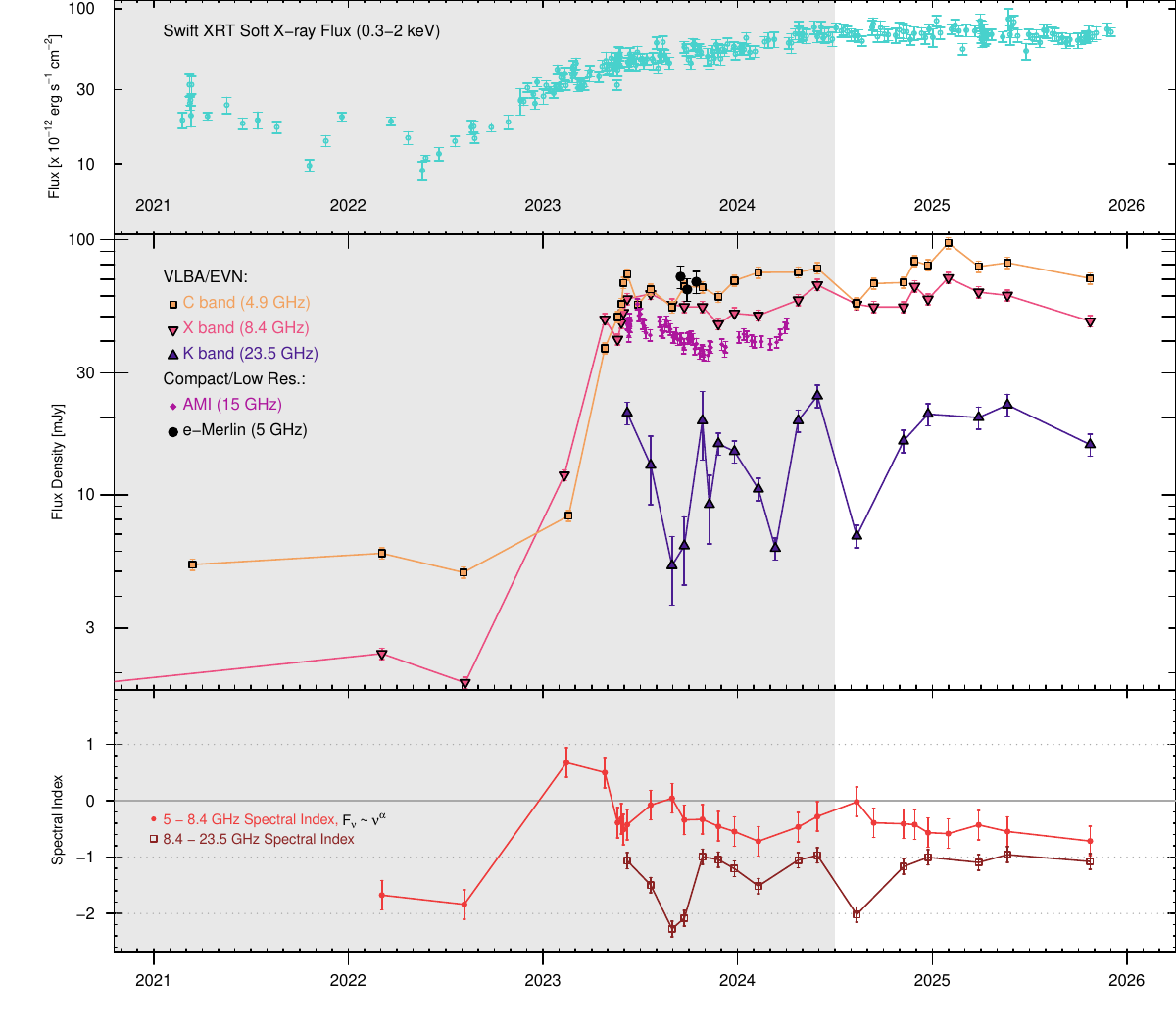}
    \caption{ Soft X-ray and radio lightcurves of 1ES\,1927+654 from 2021
    to 2026. \textit{Top:} 0.3--2~keV flux (units of
    $10^{-12}$~erg~cm$^{-2}$~s$^{-1}$) from \emph{Swift}/XRT
    \citep{laha2025}. \textit{Middle:} VLBA/EVN flux densities at C, X, and K bands (5, 8.4, and 22/23.6~GHz), along with lower-angular-resolution AMI (15.5~GHz) and e-MERLIN (5~GHz) measurements. All plotted values are integrated flux densities; the VLBA/EVN points are core-dominated at these frequencies, since no significant extended emission is detected on VLBI scales. \textit{Bottom:} Radio spectral index between 5--8.4~GHz (light red) and 8.4--23.6~GHz (dark red). The gray shaded region indicates the time period covered by previously published data \citep{meyer2025}, while data points outside this region are newly presented in this work. The radio emission remained quiescent until an exponential rise in early 2023, after which the flux stabilized near its June 2023 peak across C and X bands. New K-band epochs presented in this work are consistent with this plateau, with only modest flux variations.}
    \label{fig:lightcurve}
\end{figure*}

\subsection{VLBA Observations, Calibration, and Imaging}
\label{sec:vlba_cal}

The VLBA observations analyzed in this work were obtained between August 2024 and October 2025 as part of several approved monitoring programs. The observations provide multi-frequency coverage in the L, C, X, and K bands and were designed to trace the structural and spectral evolution of compact radio emission on parsec scales. A summary of the VLBA observations and final image properties is presented in Table~\ref{tab:Observation}, including observing date, frequency, image noise, peak and integrated flux densities, restoring beam parameters, and the degree of self-calibration applied (`no' = none; `p' = phase-only; `ap' = amplitude and phase).

All VLBA datasets were calibrated using the National Radio Astronomy Observatory Astronomical Image Processing System (AIPS; \citealt{Van_1996ASPC..101...37V}), using the development release 31DEC23. Each frequency band was calibrated independently. Standard VLBA calibration procedures were followed using tasks within the \textsc{VLBAUTIL} package, including correction for instrumental delays, bandpass calibration, amplitude calibration using system temperatures and gain curves, and phase calibration. The nearby compact source J1933+6540 was used as the phase-referencing calibrator for all observations. Bad data were flagged as necessary based on visibility inspection and calibration diagnostics. The calibrated visibilities were split using the AIPS task \textsc{split} and imaged using the \textsc{imagr} task. Imaging was performed using natural weighting to maximize sensitivity to low-surface-brightness emission. For observations with sufficient signal-to-noise ratio per antenna on short solution intervals, one or more iterations of phase-only self-calibration were applied. In several C- and X-band epochs, an additional round of amplitude and phase self-calibration was performed. Self-calibration was applied only when it resulted in a demonstrable improvement in the image root-mean-square (RMS) noise and dynamic range. The final images reported in Table~\ref{tab:Observation} correspond to the self-calibrated images where applicable, and the degree of self-calibration applied is indicated in the table.

For image analysis and flux density measurements, we additionally used the Common Astronomy Software Applications package (\textsc{CASA}; \citealt{Casa_2022PASP..134k4501C}). Integrated flux densities were measured using two-dimensional Gaussian fitting within the \textsc{viewer} tool. Image noise levels were determined from source-free regions in each image. The resulting peak and integrated flux densities are reported in Table~\ref{tab:Observation}. In the K-band VLBA observations, residual emission was apparent after initial CLEAN deconvolution of the compact core, suggesting the presence of newly emerging resolved components. To investigate this structure, we produced core-subtracted datasets following a procedure similar to~\citet{meyer2025}. Beginning from datasets with one round of phase-only self-calibration applied, we constructed a point-source model of the unresolved core using a restricted CLEAN region centered on the core position. This model was subtracted from the calibrated visibilities using the CASA task \texttt{uvsub}, after which the residual datasets were re-imaged.

\begin{figure*}[t]
    \centering
    \includegraphics[width=1\textwidth]{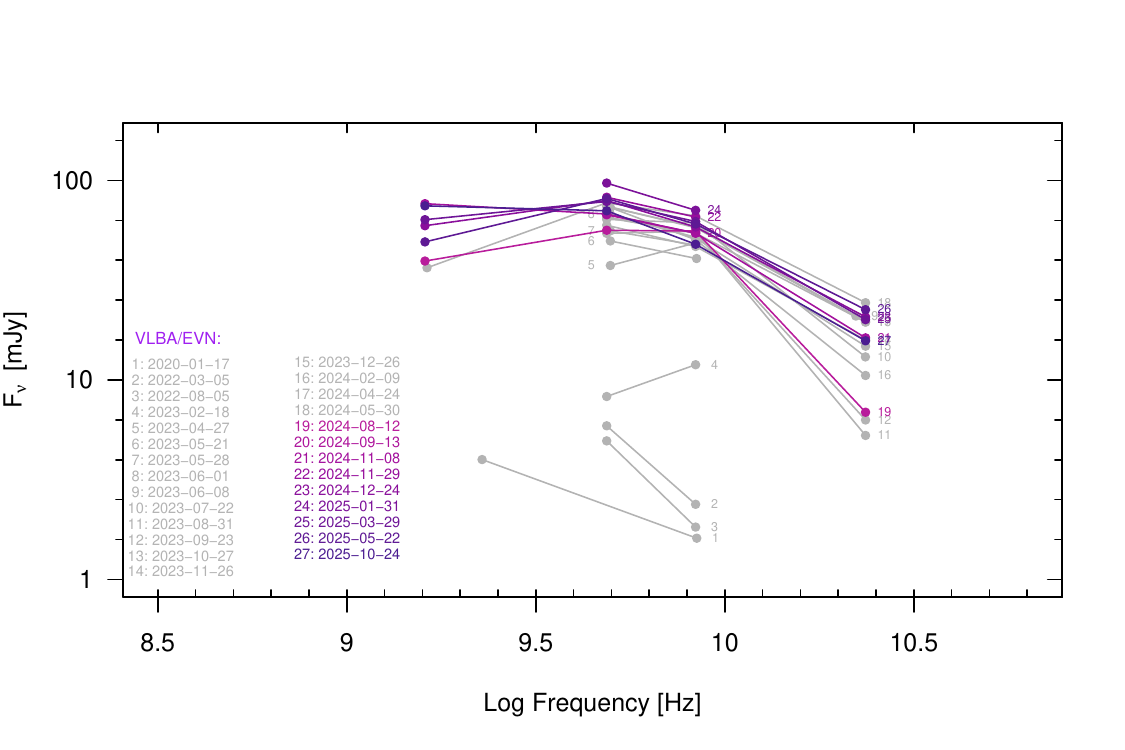}
    \caption{ VLBI radio spectral energy distribution (SED) of 1ES\,1927+654 from 2020 January through 2025 October. The previously published VLBI epochs (2020--2024 May) are displayed in gray, while the newly added VLBA observations from 2024 August through 2025 October are shown in color. The colored VLBA epochs reveal that the source exhibited a relatively stable GHz-peaked form. The spectral peak lies near (or just below) $\sim$5\,GHz and shows modest evolution across the monitoring period.}
    \label{fig:SED_vlba_only}
\end{figure*}

\subsection{VLA Multi-band Observations and Polarization} \label{sec:vla_multi}

A single-epoch, simultaneous multi-band VLA observation was obtained in August 2024 to measure a complete radio spectrum of 1ES\,1927+654. The observations covered frequencies from 1--45~GHz (L through Q bands), with an additional 340~MHz data point provided simultaneously by VLITE,
providing continuous spectral coverage from below the spectral turnover
to the optically thin regime. These data are critical for constraining the broadband radio spectral energy distribution and for linking the low-frequency VLBA measurements to higher-frequency emission associated with the compact jet and corona. All VLA data were calibrated and imaged using the \textsc{CASA} software package. Initial calibration was performed using a modified version of the standard VLA calibration pipeline, followed by additional manual flagging to mitigate radio frequency interference (RFI), particularly at lower frequencies. Polarization calibration was carried out for all bands, including corrections for cross-hand delays, instrumental polarization (D-terms), and absolute polarization position angle using standard procedures and reference values provided by NRAO. The commensal VLITE data at 340~MHz were processed separately using the standard VLITE calibration and imaging pipeline \citep{Polisensky2016}, which uses tasks from the Obit \citep{Cotton2008} and AIPS software packages. A separate Stokes~$I$ image was made for data recorded during observations at each of the eight primary observing bands using the Obit task \textsc{MFImage}. The source in each image was fit using~\textsc{PyBDSF}, and the resulting flux densities were averaged to obtain a final measurement of $37.4 \pm 4.9$~mJy at 340~MHz.

Imaging was performed using the \texttt{tclean} task with multi-term multi-frequency synthesis (\texttt{mtmfs}) and natural weighting to maximize sensitivity. Final full-Stokes (I, Q, U, V) images were produced for all bands with reliable polarization calibration. Flux densities were measured from the Stokes~I images, and polarization properties were derived from the Stokes~Q and U images. The fractional linear polarization and EVPA were computed from the polarized intensity and polarization angle maps. We note that $P = \sqrt{Q^2 + U^2}$ is positive-definite, so noise in $Q$ and $U$ adds in quadrature and biases $P$ high at low signal-to-noise; at the signal-to-noise of these measurements the correction is negligible compared to the flux calibration uncertainty. No significant extended emission was detected in any band, and the source remains unresolved at VLA angular resolution, indicating that the emission is dominated by a compact core.

\subsection{VLA X-band Monitoring} \label{sec:vla_x}

High-cadence VLA monitoring observations were carried out at X band (10~GHz) between March 2025 and July 2026. These observations provided approximately weekly to monthly sampling over about seventeen months and were designed to track the temporal evolution of the total flux density, fractional polarization, and EVPA of the compact radio emission.

All monitoring observations were calibrated using standard procedures within \textsc{CASA}, following the same calibration strategy as described in Section~\ref{sec:vla_multi}. Imaging was performed using \texttt{tclean} with natural weighting, and self-calibration was applied where justified by the data quality. Full-Stokes (I, Q, U, V) images were produced for every epoch. The source is unresolved at all epochs, so we located the peak pixel in Stokes $I$ and read $I$, $Q$, $U$ and $V$ at that same pixel, ensuring that all four Stokes parameters refer to the same position. Noise estimates were taken from source-free regions of the individual $Q$, $U$ and $V$ planes rather than from the polarized intensity image, since the latter follows a Rayleigh rather than a Gaussian distribution. We adopt systematic floors of $0.2\%$ on linear polarization, $0.3\%$ on circular polarization, and $2^\circ$ on EVPA, reflecting residual leakage and absolute angle calibration; these dominate the uncertainty budget at the signal-to-noise of these observations.

\begin{figure*}
    \centering
    \includegraphics[width=0.41\textwidth,
    clip, trim=0cm 0 0 0]{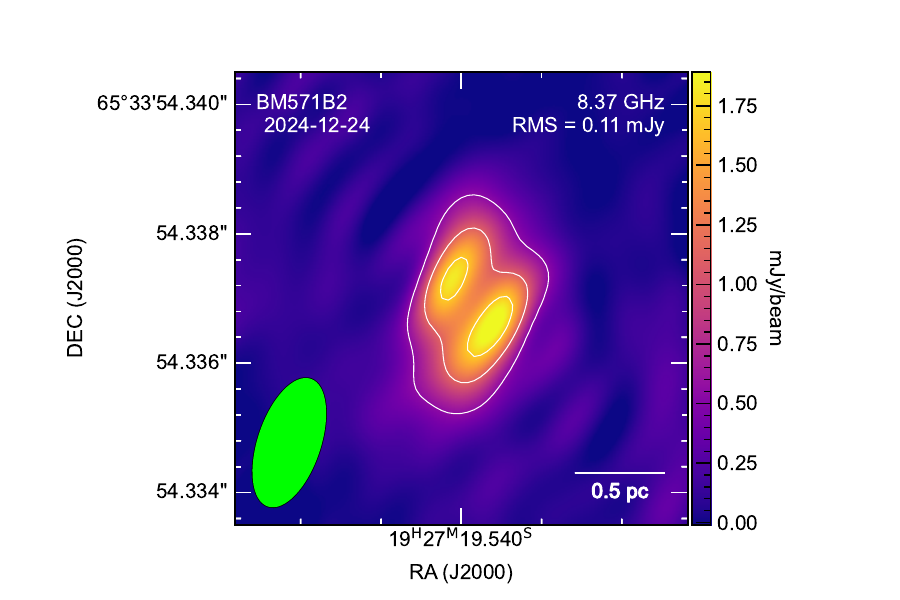}
    \hspace{-1cm}
    \includegraphics[width=0.33\textwidth, clip, trim=3cm 0 0 0]{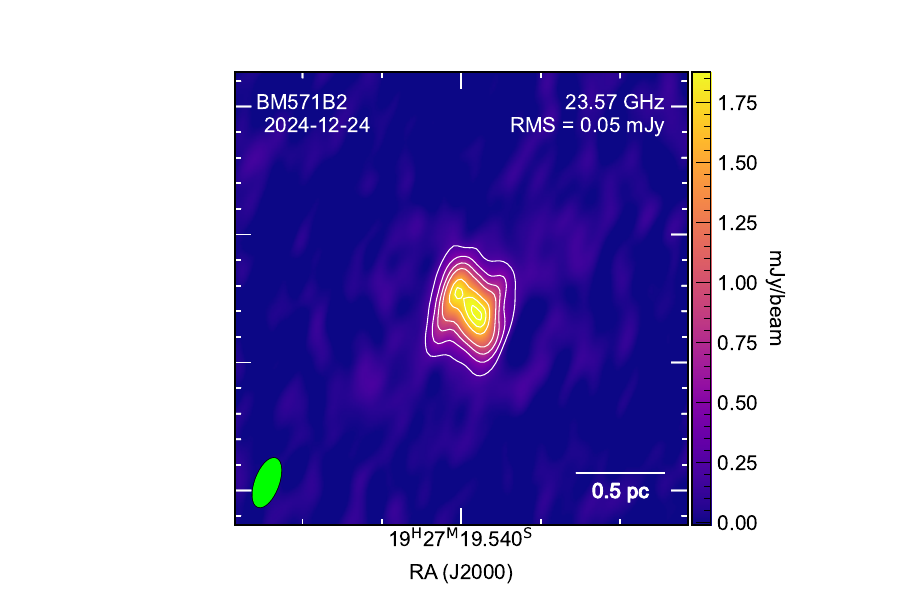}
    \hspace{-1cm}
    \includegraphics[width=0.33\textwidth, clip, trim=3cm 0 0 0]{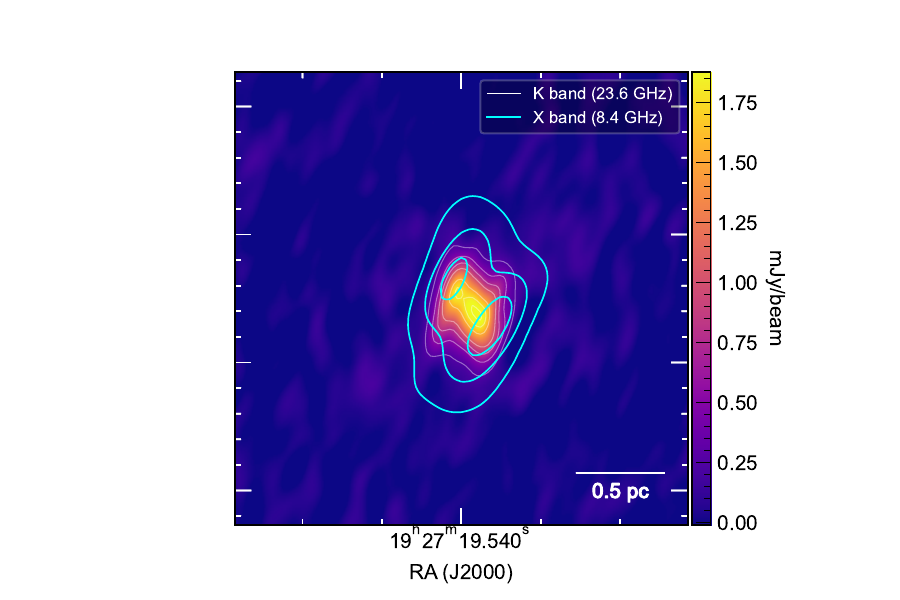}
    \caption{Core-subtracted VLBA residual images of 1ES\,1927+654 from the 2024 December epoch. \textit{Left:} X-band (8.4~GHz) residual image. \textit{Middle:} K-band (23.6~GHz) residual image. \textit{Right:} K-band image with X-band contours (cyan) overlaid. Contours are drawn at $(5,10,15,20,30,35)\times\sigma$ of the corresponding core-subtracted images. While the bipolar two-component structure was previously detected only at K band \citep{meyer2025}, the X-band residual map presented here reveals, for the first time, the same bipolar structure at a larger angular separation, consistent with outward propagation of the jet components. In addition, the higher-resolution K-band image reveals a bridge of residual emission between the two X-band peaks, possibly connecting the two components (see Section~\ref{sec:VLBA_x_band_blob}).}
    \label{fig:Xband_residual}
\end{figure*}

\section{Results and Discussion} \label{sec:results}

\subsection{VLBA}
\subsubsection{Light Curve and SED}\label{sec:VLBA_lightcurve}

Figure~\ref{fig:lightcurve} shows the long-term radio evolution of
1ES\,1927+654 from 2021–2025, combining our new VLBA monitoring (August 2024 – October 2025) with previously published VLBA, EVN, AMI, and e-MERLIN measurements. The quiescent radio state persisted through early 2023, followed by the well-documented rapid rise between February and July 2023. During this period, the integrated flux density increased by more than an order of magnitude, reaching $\sim70$~mJy at 5~GHz. 

Our newly added VLBA epochs show that, after the mid-2023 peak, the radio flux has remained remarkably stable across C, X, and K bands, with only modest ($\lesssim 10\%$) fluctuations about a plateau level. This persistent plateau indicates that the compact radio emission has not undergone further dramatic brightening since the 2023 outburst, in contrast to the rapid order-of-magnitude rise observed during the initial radio flare. Such sustained radio plateaus following an initial CL-AGN-triggered outburst have recently been highlighted in population studies of radio-emitting changing-look AGN, where the duration and stability of the post-outburst radio emission vary considerably across the known sample \citep{Birmingham2025arXiv}.

Figure~\ref{fig:SED_vlba_only} shows the VLBA radio spectral energy distribution (SED) of 1ES\,1927+654. Previously published VLBA epochs (2020–2024 May) are shown in gray, while the newly added VLBA observations (2024 August–2025 October) are shown in color. The colored epochs indicate that the source maintained a relatively stable GHz-peaked spectrum, with the spectral peak near $\sim$5~GHz and modest evolution over the monitoring period.

\subsubsection{Resolved Structure in K and X Band}\label{sec:VLBA_x_band_blob} Previous core-subtracted K-band VLBA images obtained in February 2024 reveal a persistent two-component structure, consisting of two compact peaks located on either side of the radio core. This morphology was absent during the early stages of the radio outburst in mid-2023, when the source appeared unresolved and point-like at all frequencies~\citep{meyer2025}. The appearance of this 
two-component structure marks the onset of spatially resolved emission from a nascent parsec-scale jet.

Starting in the August 2024 epoch, a similar morphology becomes detectable in the X-band (8.4~GHz) VLBA observations. The X-band images reveal two compact features symmetrically offset from the central core position, with an angular separation of approximately $\sim0.8$~mas, roughly twice the $\sim0.4$~mas separation originally observed in the K-band images. The position angle of the structure is consistent with that seen at K band, suggesting that both bands are tracing the same expanding jet system.

Figure~\ref{fig:Xband_residual} shows the core-subtracted VLBA residual images for the 2024 December epoch, with the X-band image in the left panel, the K-band image in the middle panel, and the K-band image with overlaid X-band contours (cyan) in the right panel. The bipolar structure discussed above is clearly visible in the X-band image, while the higher-frequency K-band observation provides improved angular resolution and reveals a bridge of residual emission connecting the two compact peaks. The overlay panel confirms that both bands trace the same bipolar jet axis at a consistent position angle. The presence of this bridge suggests that the two components may be part of a single continuous outflow structure rather than discrete, independently ejected knots, though it could alternatively reflect a recollimation feature or internal shock within the young relativistic jet, similar to bridges of emission connecting standing shocks to the core reported in other AGN jets~\citep{marscher2008}. The radio spectrum exhibits a turnover at $\sim3$--$5$~GHz, indicating that the emission at both X band (8.4~GHz) and K band (23.6~GHz) lies in the optically thin regime of the synchrotron spectrum. Therefore the separation observed in the X-band images is unlikely to be driven solely by synchrotron self-absorption opacity effects. Instead, it may reflect the detection of emission from plasma farther downstream of the central engine in the expanding jet flow.

Importantly, the emergence of the resolved structure at X band is not driven by changes in observational sensitivity. Prior to May 2024 the on-source integration time was approximately $\sim55$ minutes per observation, while subsequent observations used $\sim24$ minutes. Although this reduction affects the noise level, it does not alter the intrinsic source morphology. The 
synthesized X-band beams were approximately $2 \times 0.9$ mas throughout the monitoring campaign, indicating that the earlier observations had sufficient angular resolution to detect such compact structures if they had been present.

Across both X and K bands and all epochs in our monitoring campaign, the resolved structures trace the same emerging jet axis, with position angles consistently falling within $\sim35^\circ$--$45^\circ$, consistent with earlier results reported in~\citet{meyer2025}. The stability of this 
position angle across both frequency bands and the full $\sim$1.5-year monitoring baseline provides a robust estimate of the orientation of the emerging radio jet axis. A detailed epoch-by-epoch structural analysis of the VLBA data will be presented in a future study.

\begin{figure*}[t]
    \centering
    \includegraphics[width=\textwidth]{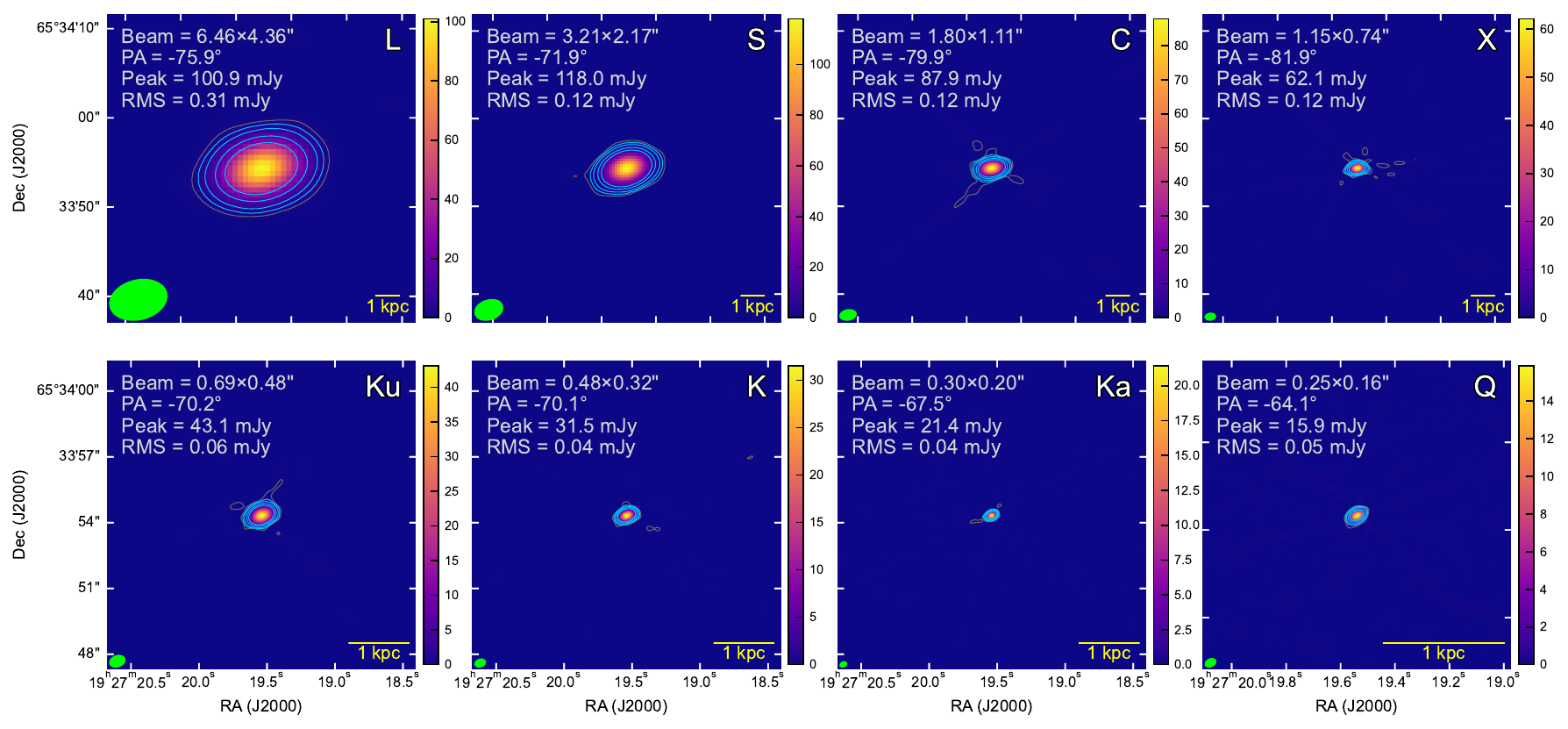}
    \caption{VLA multi-band continuum maps of 1ES\,1927+654 from L (1.5~GHz) to Q band (44~GHz). The images are aligned to the core position, with contours plotted at levels of $(5,\,10,\,15,\,20,\,50,\,100,\,\text{and}\,200)\times\sigma$ in each map, where $\sigma$ denotes the RMS noise. The filled green circle in the lower-left corner of each panel indicates the synthesized beam. The linear scale bar in the lower-right corner corresponds to 1~kpc at the source distance (74.2~Mpc). The L--X band panels show a common field of view, while the higher-frequency Ku, K, Ka, and Q bands are displayed over progressively smaller spatial scales to emphasize compact emission near the nucleus.}
    \label{fig:VLA_multiband}
\end{figure*}

\subsection{VLA Multiband Observations}
\subsubsection{Broadband Spectrum}\label{sec:vla_multiband_SED}
The VLA multiband observations spanning 1--45~GHz were obtained in August 2024, during the prolonged high-flux plateau phase of the radio outburst (see the central panel of Figure~\ref{fig:lightcurve}), which followed the exponential rise observed in early 2023. Continuum images were produced in all standard VLA bands from L (1.5~GHz) through Q (44~GHz), as shown in Figure~\ref{fig:VLA_multiband}. The corresponding image properties, including observing frequency, synthesized beam size and position angle, peak flux density, and rms noise level, are indicated directly on the images.

\begin{figure*}
    \centering
    \includegraphics[width=0.8\textwidth]{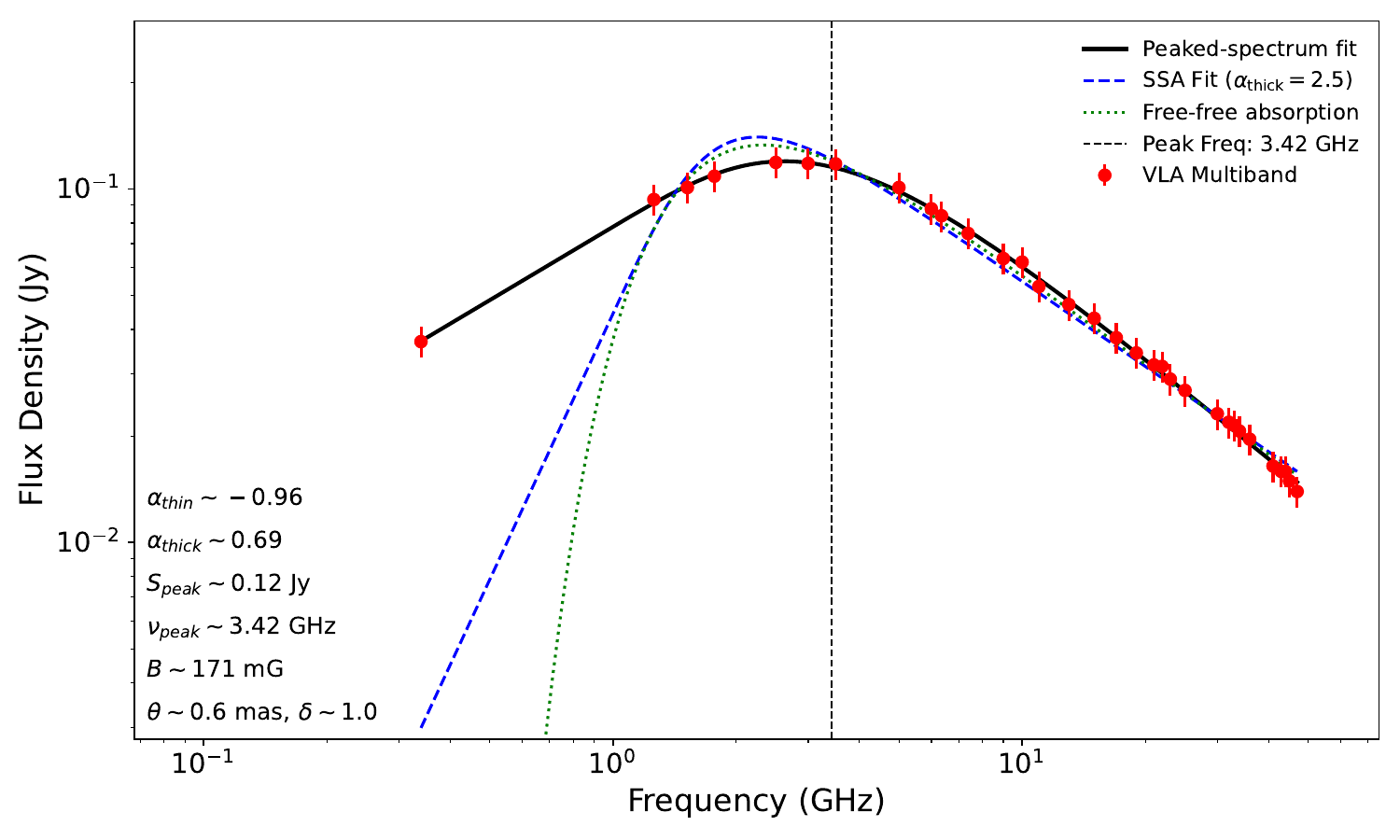}
    \caption{Radio SED of 1ES\,1927+654 with a phenomenological peaked-spectrum model, a synchrotron self-absorbed (SSA) model (with a fixed  $\alpha_{\rm thick} = 2.5$), and a free-free absorption model fit to the VLA+VLITE flux densities. The lowest-frequency data point at 340~MHz was obtained with VLITE. The vertical dashed line indicates $\nu_{\rm peak}$ as defined by the peaked-spectrum functional form (Equation~\ref{eq:peaked_form}); note that this is a shape parameter rather than the literal frequency of peak flux (see footnote~\ref{fn:nupeak}). Fit parameters ($S_{\rm peak}$, $\nu_{\rm peak}$, $\alpha_{\rm thick}$, $\alpha_{\rm thin}$), the derived magnetic field $B$, assumed VLBI core size $\theta$, and Doppler factor $\delta$ are annotated in the figure.}
    \label{fig:ssa_fit}
\end{figure*}

The measured flux densities of the unresolved nuclear component, including the 340~MHz VLITE point, exhibit a convex radio spectrum characteristic of Gigahertz-Peaked Spectrum (GPS) sources, with a well-defined spectral turnover near $\sim$3~GHz (Figure~\ref{fig:ssa_fit}). Such peaked spectra are commonly interpreted as arising from synchrotron emission produced in compact, optically thick radio sources, where absorption processes dominate at low frequencies \citep[e.g.,][]{odea1991,odea1998,Snellen_1998A&AS..131..435S}. To parameterize the broadband radio spectrum, we fit a phenomenological peaked-spectrum model that has been widely applied to GPS and compact steep-spectrum (CSS) sources in the literature \citep[e.g.,][]{Snellen_1998A&AS..131..435S,Kim_2022MNRAS.510..815K}. The adopted functional form is

\begin{equation}
\label{eq:peaked_form}
\begin{split}
S(\nu) &= \frac{S_{\rm peak}}{1 - e^{-1}} 
         \left( \frac{\nu}{\nu_{\rm peak}} \right)^{\alpha_{\rm thick}} \\
       &\quad \times \left[ 1 - \exp \left( - \left( \frac{\nu}{\nu_{\rm peak}} \right)^{\alpha_{\rm thin} - \alpha_{\rm thick}} \right) \right],
\end{split}
\end{equation}

where $S_{\rm peak}$ and $\nu_{\rm peak}$ denote the peak flux density and turnover frequency, respectively\footnote{\label{fn:nupeak}$\nu_{\rm peak}$ is a shape/normalization parameter of the functional form rather than the literal frequency of maximum flux density. Due to the asymmetry between $\alpha_{\rm thick}$ and $\alpha_{\rm thin}$, the true flux maximum occurs at a frequency slightly below $\nu_{\rm peak}$, as visible in Figure~\ref{fig:ssa_fit}.}, while $\alpha_{\rm thick}$ and $\alpha_{\rm thin}$ describe the optically thick and optically thin spectral indices, both of which are free parameters of the fit rather than physically imposed values. Fitting the VLA+VLITE flux densities and adopting a conservative 10\% systematic uncertainty on the flux densities, we obtain best-fit parameters of $S_{\rm peak} = 0.12 \pm 0.01$~Jy, $\nu_{\rm peak} = 3.42 \pm 0.35$~GHz, $\alpha_{\rm thick} = 0.69 \pm 0.08$, and $\alpha_{\rm thin} = -0.96 \pm 0.04$.

The rising portion of the spectrum is therefore significantly flatter than the canonical $\alpha_{\rm thick} = 2.5$ expected for a homogeneous, single-zone synchrotron source, differing from that value by $\sim23\sigma$. To test whether this flattening indicates an additional absorption mechanism, we refit the spectrum with two alternatives. Fixing $\alpha_{\rm thick} = 5/2$ raises $\chi^2$ from 2.0 (28 dof) to 106 (29 dof) and underpredicts the 340~MHz VLITE point by $9.2\sigma$. A homogeneous free--free absorption model, $S(\nu) = S_0\,\nu^{\alpha_{\rm thin}}\exp\!\left(-\tau_{\rm ff}\,\nu^{-2.1}\right)$ \citep[e.g.,][]{Kameno2003, Orienti2008, Shao2022}, gives $\chi^2 = 112$ for 29 degrees of freedom and underpredicts the same point by $10\sigma$. Free--free absorption in a homogeneous screen attenuates the spectrum exponentially below the turnover, producing a steeper low-frequency cutoff than synchrotron self-absorption; this is the behavior disfavored by the fit above. We note, however, that the asymptotic slope is only approached well below the turnover, and that close to $\tau \sim 1$ the spectrum can appear flatter. Moreover, free--free absorption in an internal or inhomogeneous external screen predicts optically thick indices of $\alpha_{\rm thin} + 2.1$ and $\alpha_{\rm thin} + 2.1(p+1)$ respectively \citep{Shao2022}, both of which can fall well below $5/2$. Our measurement therefore rules out homogeneous synchrotron self-absorption, but does not exclude free--free absorption in general. Such deviations are commonly observed in GPS sources and are generally attributed to source inhomogeneity, multiple unresolved emitting components, or gradients in magnetic field strength and particle density along the jet \citep{odea1998,murgia02,callingham2017}. We therefore attribute the departure of $\alpha_{\rm thick}$ from $5/2$ to inhomogeneity, most plausibly in the synchrotron-emitting region.

We note that the VLA and VLBA flux densities are consistent above 8~GHz, confirming that the radio emission at these frequencies is dominated by a compact core with negligible contribution from extended emission. At lower frequencies, however, the VLA flux density slightly exceeds the VLBA measurement, indicating a modest contribution from extended emission that is resolved out on VLBA baselines. This accounts for the slight difference in apparent spectral peak frequency between the VLA and VLBA-only SEDs 
(Figures~\ref{fig:SED_vlba_only} and \ref{fig:ssa_fit}). 

The presence of a spectrally stable, long-lived GHz-frequency turnover across multiple epochs, together with the compact, unresolved morphology of the radio core, indicates that the radio emission is dominated by synchrotron radiation from a compact, self-absorbed jet. Within the standard GPS framework, such sources are widely interpreted as young radio jets with characteristic linear sizes $\lesssim 1$~kpc, observed during the early stages of their expansion \citep[e.g.,][]{odea1998,snellen2000,Cheng2023}. In this context, the observed spectral curvature supports the interpretation that 1ES\,1927+654 currently hosts a newly formed, compact radio jet. While free--free absorption may contribute to the low-frequency turnover in some GPS sources \citep[e.g.,][]{Begelman1999, Stawarz2008}, the homogeneous model comparison above does not favor it in 1ES\,1927+654, and we interpret the spectrum as synchrotron self-absorption in an inhomogeneous source, while noting that an inhomogeneous free--free contribution cannot be excluded.

\subsubsection{Multiband Polarization}\label{sec:vla_multiband_polarization}
\begin{figure*}
    \centering
    \includegraphics[width=0.8\textwidth]{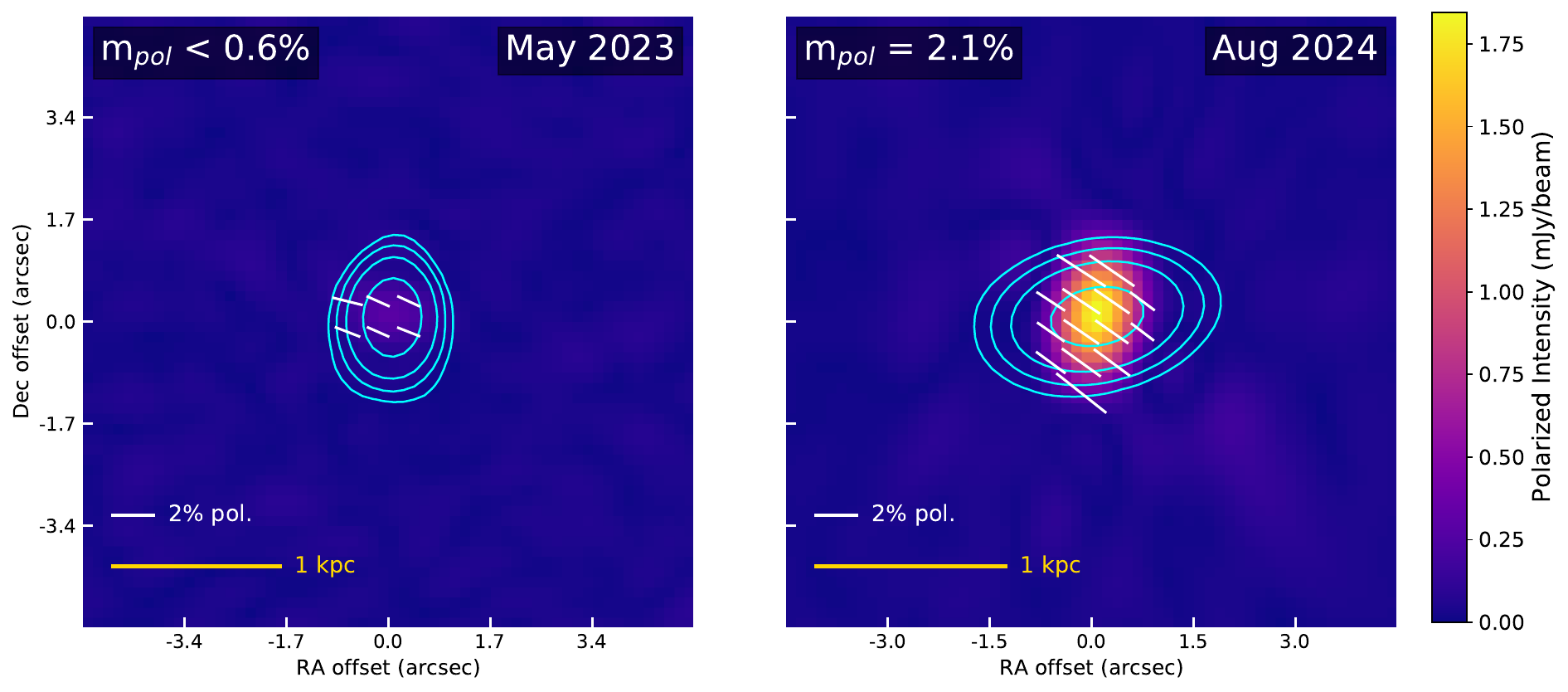}
    \caption{ VLA C-band linear polarization images of 1ES\,1927+654 from (left) May~2023~\citep[shown before in][]{meyer2025} and (right) August~2024. Colors show the linearly polarized intensity in mJy~beam$^{-1}$. Contours trace the total intensity (Stokes~$I$), with rms values of 0.055~mJy~beam$^{-1}$ (May~2023) and 0.12~mJy~beam$^{-1}$ (August~2024). White line segments indicate the projected magnetic-field orientation, obtained by rotating the electric vector position angles by $90^{\circ}$. The color scale is identical in both panels to enable a direct comparison of polarized intensity between epochs. The May~2023 image shows only marginal or undetected polarization (upper limits), whereas the August~2024 image reveals a fractional polarization of 2.1\% and a more ordered magnetic-field structure.}
    \label{fig:c_band_pol}

\end{figure*}

\begin{figure*}[htbp]
    \centering
    \includegraphics[width=1\textwidth]{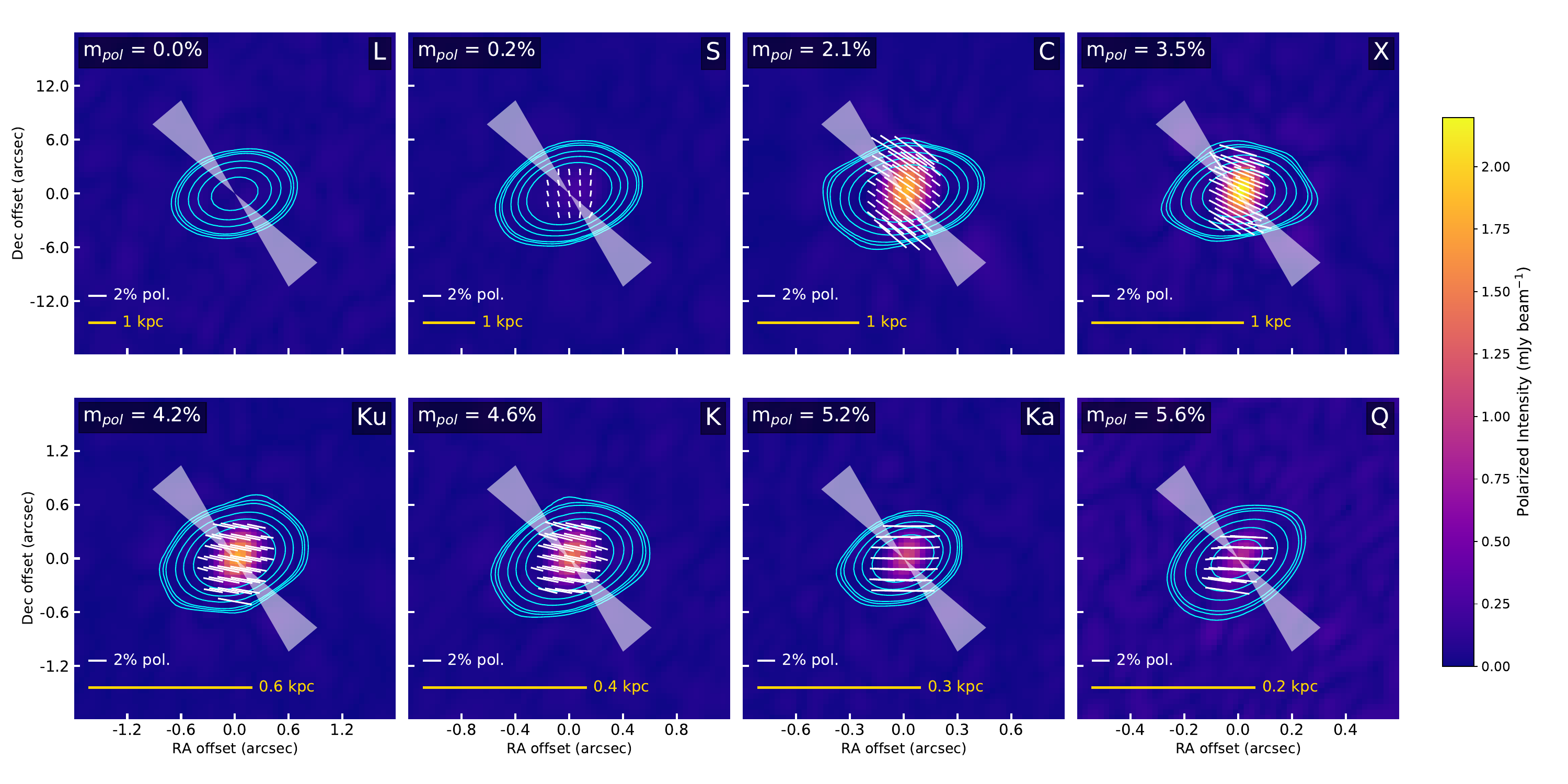}
    \caption{VLA Multi-band linear polarization images of 1ES\,1927+654. The background color shows the polarized intensity (mJy/beam) at each band: L (1.5~GHz), S (3~GHz), C (6~GHz), X (10~GHz), Ku (15~GHz), K (22~GHz), Ka (33~GHz), and Q (44~GHz), while cyan contours represent the total intensity from the corresponding VLA band. Magnetic field vectors, derived by rotating the EVPA by 90° to trace the projected magnetic field orientation, are overlaid and averaged in 5$\times$5 pixel boxes. The white shaded bipolar cones indicate the projected jet/outflow direction, centered at a position angle of $40^\circ$ (measured from north to east) as measured from the extended K-band VLBA emission (see section~\ref{sec:VLBA_x_band_blob}), with a $\pm10^\circ$ opening angle shown to illustrate the plausible range of the outflow orientation. The cones are drawn with an exaggerated spatial extent for clarity, as no resolved large-scale outflow is detected in the VLA images; they are intended as a schematic guide to the jet angle and magnetic field orientation inferred from the VLBA observations. The polarization fraction at the central peak is indicated in the top-left of each panel. The polarization increases with frequency, consistent with a transition from depolarized emission in the outer, turbulent sheath of the jet to optically thin emission from the inner, more ordered jet/core region.
    }
    \label{fig:vla_multiband_pol}
\end{figure*}

VLA multiband observations of 1ES\,1927+654 provide the first robust detection and characterization of its radio polarization properties across the S--Q bands. Full polarization calibration was performed, enabling detailed measurements of the frequency-dependent fractional polarization, Faraday rotation, and EVPA behavior. Importantly, no significant polarization was detected in the 2023 C-band VLA data, where the fractional polarization was constrained to $\lesssim0.6\%$~\citep{meyer2025}. In contrast, the 2024 observations reveal clear polarization detections at C ($\sim2.1\%$) band (Figure~\ref{fig:c_band_pol}) and increasing steadily toward higher frequencies, marking the first emergence of detectable polarized emission from this source (Figure~\ref{fig:vla_multiband_pol}). Table~\ref{tab:evpa_lambda2} lists all the measured VLA multi-band polarization properties.

We measured the linear fractional polarization ($m_{pol}$) for each band as

\begin{equation}
m_{pol} = \frac{P}{I} \times 100\%,
\end{equation}

where $P$ is the linearly polarized intensity and $I$ is the total intensity.  
Both $P$ and $I$ were measured using the central pixel of the source in the respective Stokes $Q/U$ and $I$ images. The uncertainty in the fractional polarization, $\sigma_m$, was calculated by propagating the errors in $P$ and $I$:

\begin{equation}
\sigma_m = m \sqrt{ \left(\frac{\sigma_P}{P}\right)^2 + \left(\frac{\sigma_I}{I}\right)^2 } ,
\end{equation}

where $\sigma_P$ is the RMS noise measured in a source-free corner of the polarization image, and $\sigma_I$ is the total uncertainty in the total intensity, which includes both the thermal noise and a conservative flux calibration uncertainty:

\begin{equation}
\sigma_I = \sqrt{ (\mathrm{rms}_I)^2 + (f_{\rm cal}\, I)^2 }.
\end{equation}

Here, $\mathrm{rms}_I$ is the RMS measured in a source-free corner of the Stokes $I$ image, and $f_{\rm cal} = 0.05$ corresponds to a 5\% calibration uncertainty typical for VLA observations.

The $m_{pol}$ increases systematically from $\sim0.2\%$ at $\sim3$~GHz (S band) to $\sim3.5\%$ at 10~GHz (X band), reaching $\sim6\%$ at 44~GHz (Q band). This roughly 30-fold increase in fractional polarization with frequency is consistent with the general expectation of Faraday depolarization, in which polarized emission at longer wavelengths is reduced by differential Faraday rotation arising from magnetized plasma either internal to or surrounding the emitting region \citep{burn1966,tribble1991}.

At low frequencies, optical depth effects near the synchrotron self-absorption turnover can blend emission from multiple unresolved regions with differing magnetic field orientations, suppressing the net observed polarization independently of any Faraday rotation \citep{pacholczyk1970, Jones1977}. Separately, wideband VLA observations of other AGN have shown that sources with polarization below instrumental detection limits ($\lesssim 1\%$) at L band (1.4 GHz) can exhibit polarization of several percent at higher frequencies once these opacity effects, together with Faraday depolarization from an intervening screen, become negligible \citep{pasetto2016}. The non-detection of polarization in the 2023 C-band data ($m_{\rm pol} \lesssim 0.6\%$) followed by clear detections across all bands in 2024 may therefore reflect the combined effect of these two mechanisms becoming less significant as the source evolved, equivalent to the effective disappearance or weakening of an external Faraday screen as the optically thin jet emission came to dominate.

The high fractional polarization detected in the K, Ka, and Q bands is therefore consistent with synchrotron emission arising in an optically thin jet with an ordered magnetic field. The corresponding EVPA values remain within $\sim10^\circ$ of each other across the K, Ka, and Q bands (Table~\ref{tab:evpa_lambda2}), indicating a spectrally stable magnetic-field orientation at high frequencies within this epoch and that Faraday depolarization and opacity effects are minimal in this regime. The inferred magnetic-field position angles are not found to be strictly parallel to the jet axis defined by the VLBA K-band two-component structure. They instead suggest an ordered magnetic field whose orientation is offset from the jet axis, potentially shaped by oblique shocks, shear, or evolving jet collimation near the launch region. Such a behavior is commonly observed in polarimetric studies of young or newly reactivated AGN jets, where shock compression of turbulent plasma can produce partially ordered magnetic fields with orientations that deviate from simple configurations aligned along or perpendicular to the jet axis \citep{laing1980,hughes1985,marscher2008,silpa2021}. Together, these results indicate that the polarized emission in 1ES\,1927+654 (Figure~\ref{fig:vla_multiband_pol}) traces the emergence of an ordered magnetic field whose orientation departs from the jet axis during the early stages of jet formation --- the first such detection in this source.

\subsubsection{EVPA, Rotation Measure and Faraday Structure}\label{sec:vla_evpa}

The electric vector position angle (EVPA), $\chi$, describes the orientation of the linear polarization with respect to north, measured through east, and is defined as
\begin{equation}
\chi = \frac{1}{2} \arctan\left(\frac{U}{Q}\right),
\end{equation}
where $Q$ and $U$ are the Stokes parameters of linear polarization. The EVPA uncertainty was estimated as
\begin{equation}
\sigma_\chi \approx 28.65 \, \frac{\sigma_P}{P} \; \mathrm{deg},
\end{equation}
where $P$ is the polarized intensity and $\sigma_P$ is the rms noise measured in a source-free region of the polarization image. The factor $28.65$ arises from the conversion of the error from radians to degrees in the standard formula. 

\begin{figure}
\centering
\includegraphics[width=\columnwidth]{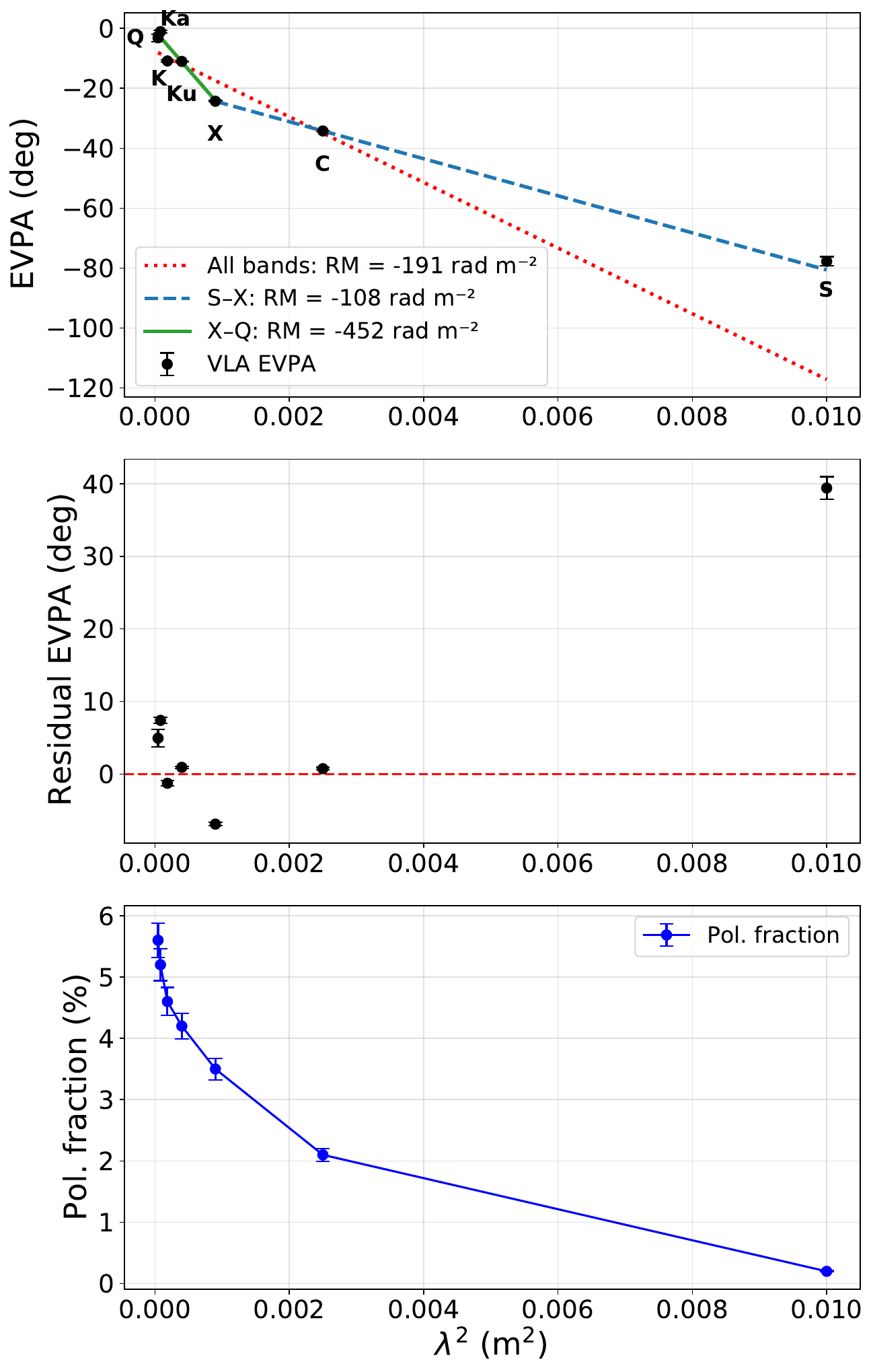}
\caption{ \textit{Top:} EVPA as a function of wavelength squared ($\lambda^2$) for VLA multi-band observations of 1ES\,1927+654. The red dotted line represents the linear fit over all bands (S--Q), weighted by the inverse-square of the EVPA uncertainties, the blue dashed line shows the fit to the low-frequency subset (S--X), and the green solid line shows the high-frequency fit (X--Q). The low-frequency rotation measure, ${\rm RM}_{\rm S-X} = -108$~rad~m$^{-2}$, likely traces an external Faraday screen or extended magnetized plasma, as suggested by slightly higher VLA flux compared to VLBA (see Section~\ref{sec:vla_multiband_SED}). The high-frequency rotation measure, ${\rm RM}_{\rm X-Q} = -452$~rad~m$^{-2}$, traces a magnetized sheath external to the emitting region (see Section~\ref{sec:vla_evpa}). The global fit across all bands (S--Q) yields ${\rm RM}_{\rm all} = -191$~rad~m$^{-2}$, but with a very large reduced chi-square ($\chi^2_\nu \approx 417$), indicating that a single RM cannot adequately describe the full frequency range. \textit{Middle:} Residuals of the measured EVPAs relative to the global S--Q fit, highlighting systematic deviations and demonstrating the frequency-dependent nature of the Faraday rotation. \textit{Bottom:} Fractional polarization as a function of $\lambda^2$ with error bars computed from propagated uncertainties in polarized and total intensity. The polarization fraction increases with frequency, consistent with depolarization at longer wavelengths due to Faraday effects.
}
\label{fig:evpa_lambda2}
\end{figure}

The top panel in Figure~\ref{fig:evpa_lambda2} shows EVPA as a function of wavelength squared for the VLA multi-band observations of 1ES\,1927+654, spanning frequencies from 3 to 44~GHz. The EVPAs were measured from the polarization angle images by averaging the Stokes $Q$ and $U$ emission within a common $5\times5$ pixel region centered on the radio core, which approximately samples one synthesized beam and maximizes signal-to-noise while minimizing beam-related biases. The EVPA uncertainties were derived from the rms noise in the polarized intensity images. Prior to fitting, the EVPAs were unwrapped to remove $n\pi$ ambiguities arising from the inherent $180^\circ$ periodicity of polarization angles \citep{Hovatta2012}.

The observed EVPAs were fit using the standard Faraday rotation relation
\begin{equation}
\chi(\lambda^2) = \chi_0 + \mathrm{RM}\,\lambda^2 ,
\end{equation}
where $\chi_0$ is the intrinsic EVPA at zero wavelength and RM is the rotation measure.

Typically a linear dependence of EVPA on $\lambda^2$ is expected for Faraday rotation produced by a uniform magnetized plasma along the line of sight \citep{burn1966}. However, deviations from linearity are commonly observed in compact AGN cores and indicate more complex Faraday structures \citep{osullivan2012,Pasetto2018}. In our case, the EVPA behavior exhibits clear deviations from a single linear $\lambda^2$ relation over the full frequency range. Motivated by this behavior, we performed three weighted linear least-squares fits, with weights given by $w_i = 1/\sigma_{\chi,i}^2$, where $\sigma_{\chi,i}$ is the EVPA uncertainty at each frequency \citep{Kravchenko2017,Pasetto2018}. Fits were carried out for three different frequency selections: (i) the full frequency range (S--Q bands), (ii) the low-frequency subset (S--X bands), and (iii) the high-frequency subset (X--Q bands). The resulting slope/rotation measures are ${\rm RM}_{\rm all} = -191~{\rm rad~m^{-2}}$, ${\rm RM}_{\rm S\text{--}X} = -108~{\rm rad~m^{-2}}$, and ${\rm RM}_{\rm X\text{--}Q} = -452~{\rm rad~m^{-2}}$, respectively. We caution that the S and C bands lie at or near the SSA turnover frequency ($\nu_{\rm peak} \approx 3.4$~GHz), where the optically thin assumption underlying the standard RM formula may not hold. The low-frequency ${\rm RM}_{\rm S\text{--}X}$ should therefore be interpreted with caution, as it may be associated with differential opacity effects rather than pure Faraday rotation. The high-frequency rotation measure (${\rm RM}_{\rm X\text{--}Q} = -452$~rad~m$^{-2}$, measured using X through Q bands) is more reliable, as X band and above are clearly in the optically thin regime. We also note that bandwidth depolarization is negligible for the high-frequency bands (X through Q), as the small $\lambda^2$ values at these frequencies result in negligible differential Faraday rotation across each spectral window.

The middle panel shows the residuals with respect to the global (S--Q) fit. The global fit yields a very large reduced chi-square, $\chi^2_\nu \approx 417$, indicating that a single rotation measure cannot adequately describe the EVPA behavior across the full frequency range. This demonstrates the presence of frequency-dependent Faraday effects, such as multiple Faraday-rotating components, beam depolarization at long wavelengths, or a stratified magnetized medium along the line of sight~\citep{Hovatta2012, osullivan2012, Pasetto2018}.

At low frequencies (S--X bands), the observed rotation measure ${\rm RM}_{\rm S\text{--}X} = -108~{\rm rad~m^{-2}}$ is relatively modest and consistent with a significant contribution from an external Faraday screen. Galactic foreground contributions are expected to be small in this direction. Using the Galactic RM maps of \citet{Hutschenreuter2022}, the Milky Way contribution at the position of 1ES\,1927+654 is of order $\sim +30$--$40~{\rm rad~m^{-2}}$, implying that the observed low-frequency RM is dominated by material local to the source, such as circumnuclear gas or extended ionized emission on VLA scales. Subtracting this Galactic foreground contribution from the observed ${\rm RM}_{\rm S\text{--}X} = -108$~rad~m$^{-2}$ yields an intrinsic source RM of approximately $-140$ to $-150$~rad~m$^{-2}$, confirming that the low-frequency RM is dominated by material intrinsic to the source rather than the Galactic foreground.

In contrast, the high-frequency X--Q band fit yields a much steeper rotation measure, ${\rm RM}_{\rm X\text{--}Q} = -452~{\rm rad~m^{-2}}$, which likely probes the immediate vicinity of the compact radio core. At these frequencies, the VLA and VLBA flux densities are consistent within the uncertainties, indicating negligible contamination from extended emission. The Faraday rotation measured in this regime therefore traces magnetized plasma in the immediate surroundings of the jet, such as a magnetized sheath or cocoon, where external Faraday screens become increasingly transparent~\citep{zavala2004,osullivan2012}. We note that a Faraday screen mixed with the synchrotron-emitting plasma would produce departures from the $\lambda^2$ law and limit the observable rotation, neither of which is seen across the X through Q bands, supporting a screen external to but closely surrounding the emitting region.

The frequency dependence of the fractional polarization provides an independent constraint on the Faraday structure inferred from the EVPA and rotation measure analysis. The fractional polarization increases rapidly from the S to X bands ($\sim0.2\%$ to $\sim3.5\%$), followed by a more gradual rise from the Ku through Q bands ($\sim4.2\%$ to $\sim5.6\%$; Table~\ref{tab:evpa_lambda2}), indicating that the strongest depolarization occurs at the lowest frequencies, while the polarization properties evolve more slowly above $\sim15$ GHz. At low frequencies, the observed polarization is strongly reduced, to $\lesssim 5\%$ of its maximum value ($m_{\rm pol} \simeq 0.2\%$ at S band, compared with $\simeq 5.6\%$ at Q band). This reduction can be naturally explained by beam depolarization due to a tangled magnetic field in the foreground or jet sheath, where the magnetic field correlation length is smaller than the observing beam~\citep[e.g.,][]{burn1966,tribble1991,Sokoloff_1998MNRAS.299..189S}. At high frequencies, the increasing fractional polarization suggests that the emission is progressively less affected by these depolarizing effects, allowing the intrinsic polarization of the synchrotron-emitting region to emerge more clearly. Combining the Faraday rotation analysis with the gigahertz-peaked synchrotron spectrum and the emerging polarization, we infer that the radio emission in 1ES\,1927+654 originates from a compact, newly formed jet embedded in a stratified, magnetized environment. The high-frequency rotation measure then reflects this inner sheath, while the stability of the EVPA across the K, Ka, and Q bands ($\sim10^\circ$; Table~\ref{tab:evpa_lambda2}) indicates an ordered field orientation in that region. The lower-frequency behavior instead reflects depolarization and Faraday rotation in a more turbulent, outer screen. This picture is consistent with early-stage jets observed in young radio AGN and in systems where jet activity has recently been triggered or reactivated \citep[e.g.,][]{odea1998,callingham2017}. We note that the observed frequency dependence of the rotation measure likely reflects a combination of beam depolarization, Faraday depth averaging, and genuine changes in the physical properties of the magnetized plasma probed at different frequencies, with higher frequencies preferentially tracing more compact regions closer to the jet base, consistent with the standard core-shift picture in which the synchrotron self-absorption photosphere recedes toward the central engine at higher frequencies \citep{Blandford1979}.

\subsubsection{Magnetic Field Strength}
Assuming that the observed Faraday rotation arises from a uniform magnetized plasma along the line of sight, the rotation measure (RM) can be expressed as
\begin{equation}
\mathrm{RM} = 812 \int n_e\, B_\parallel\, dl \ \ \mathrm{rad~m^{-2}},
\end{equation}
where $n_e$ is the electron density in cm$^{-3}$, $B_\parallel$ is the line-of-sight magnetic field component in mG, and $dl$ is the path length in parsecs \citep[e.g.,][]{burn1966, O'Sullivan_2009, Lai_2025MNRAS.540....1L}. To estimate the magnetic field, we invert the standard relation:
\begin{equation}
\langle B_\parallel \rangle \simeq \frac{|\mathrm{RM}|}{812 \, n_e \, L}.
\end{equation}

Adopting a characteristic path length $L = 0.05$~pc, consistent with parsec-scale Faraday screens in AGN jets \citep[e.g.,][]{Marti2015, O'Sullivan_2009}, and assuming plausible electron densities in the range $n_e \sim 0.1 - 10~\mathrm{cm^{-3}}$ for the inner jet or sheath plasma \citep[e.g.,][]{Hovatta2012, Kravchenko2017, Lisakov2021}, we obtain $\langle B_\parallel \rangle \sim 1 - 100~\mathrm{mG}$ 
(line-of-sight-averaged component).  We note that this estimate assumes optically thin conditions throughout; given that S and C bands lie near the SSA turnover, the low-frequency RM contribution introduces additional uncertainty, and the lower end of this range should be treated with caution. The factor-of-100 uncertainty in this estimate is dominated by the assumed range in $n_e$; future constraints from X-ray column density 
measurements or optical emission-line diagnostics could significantly narrow this range \citep[e.g.,][]{laha2022}.

We note that $\langle B_\parallel \rangle$ represents only the line-of-sight component of the total magnetic field. Since the EVPA measurements provide an estimate of the projected field orientation on the sky, the total field strength can in principle be approximated as $|B| \approx \langle B_\parallel \rangle / \cos\theta_B$, where $\theta_B$ is the angle between the field direction and the line of sight. If the field is predominantly transverse to the line of sight, as expected for a poloidal field in a jet oriented close to the plane of the sky, consistent with the two-sided parsec-scale structure and the modest apparent advance speed of $\sim0.3c$ \citep{meyer2025}, then $|B|$ could be significantly larger than $\langle B_\parallel \rangle$, potentially bringing the RM-based estimate into closer agreement with $B_{\rm SSA}$ derived below. Alternatively, for a symmetric toroidal field in a jet near the plane of the sky, the near and far sides contribute rotation measure of opposite sign, so that the net line-of-sight component largely cancels and a transverse RM gradient is produced instead; in that case $\langle B_\parallel \rangle$ would likewise underestimate the total field strength. The ratio $\sigma_{\rm RM}/|\mathrm{RM}|$ would additionally inform whether the turbulent or ordered component of the magnetic field dominates; we defer this analysis to a future study with expanded frequency coverage~\citep{burn1966}.

The inferred range $\langle B_\parallel \rangle \sim 1$--$100$~mG is in good agreement with magnetic fields inferred in the inner parsec-scale regions of AGN jets \citep{Zamaninasab2014,O'Sullivan_2009} and indicates that the Faraday-rotating medium is relatively low density compared to the narrow-line region. We caution, however, that while the path length $L \sim 0.05$~pc characterizes the coherence scale of the Faraday-rotating medium, it does not directly constrain the distance of this screen from the central engine; the screen could plausibly range from sub-parsec to several parsecs from the black hole. The extrapolation to the jet launching region therefore carries an additional geometric uncertainty beyond the factor-of-100 range in $n_e$. Under the assumption that the Faraday screen is located at $\sim1$~pc from the central engine, broadly consistent with the parsec-scale VLBA structure discussed in Section~\ref{sec:VLBA_x_band_blob}, and adopting a toroidal field scaling $B \propto r^{-1}$, the inferred fields extrapolated to $\sim10$~gravitational radii (for $M_{\rm BH} \approx 10^6\,M_\odot$) suggest magnetic field strengths of order $B \sim 10^3$--$10^5$~G near the jet launching region, where magnetically driven jet models predict the strongest fields \citep{Blandford1977, Marti2015}. This range agrees with independent estimates for this source: a minimum coronal field of $\sim10^3$~G required to sustain the observed 2--10~keV luminosity, and $B_0 \gtrsim 10^5$~G at the jet base inferred from the same $r^{-1}$ extrapolation of parsec-scale measurements \citep{Laha2025_corona, laha2025}. We further caution that if the turbulent component of the Faraday-rotating medium dominates ($\sigma_{\rm RM}/|\mathrm{RM}| \gtrsim 1$), the effective coherence length of the field would be shorter than $L$, and the inferred $\langle B_\parallel \rangle$ would represent a lower limit on the local field strength.

\begin{deluxetable*}{ccccccccccc}
\tablecaption{VLA multi-band polarization properties.\label{tab:evpa_lambda2}}
\tablehead{
\colhead{Band} & \colhead{Frequency} & \colhead{$\lambda^2$} & 
\colhead{$m_{\rm pol}$} & \colhead{$\sigma_m$} & 
\colhead{EVPA} & \colhead{$\sigma_{\rm EVPA}$} & 
\colhead{Residual} & \colhead{EVPA$_{\rm wrap}$} & \colhead{B-field} \\
\colhead{} & \colhead{(GHz)} & \colhead{(m$^2$)} & 
\colhead{(\%)} & \colhead{(\%)} & 
\colhead{(deg)} & \colhead{(deg)} & 
\colhead{(deg)} & \colhead{(deg)} & \colhead{(deg)}
}
\startdata
S  & 3  & $1.00\times10^{-2}$ & 0.20 & 0.01 & $-77.77$ & 1.56 & 39.40   & 102.15 & 12.15 \\
C  & 6  & $2.50\times10^{-3}$ & 2.10 & 0.11 & $-34.21$ & 0.18 & 0.75    & 145.81 & 55.81 \\
X  & 10 & $9.00\times10^{-4}$ & 3.50 & 0.18 & $-24.30$ & 0.22 & $-6.88$ & 155.77 & 65.77 \\
Ku & 15 & $4.00\times10^{-4}$ & 4.20 & 0.21 & $-11.01$ & 0.14 & 0.94    & 169.01 & 79.01 \\
K  & 22 & $1.86\times10^{-4}$ & 4.60 & 0.23 & $-10.85$ & 0.36 & $-1.25$ & 169.18 & 79.18 \\
Ka & 33 & $8.26\times10^{-5}$ & 5.20 & 0.26 & $-1.06$  & 0.40 & 7.41    & 178.93 & 88.93 \\
Q  & 44 & $4.65\times10^{-5}$ & 5.60 & 0.28 & $-3.12$  & 1.21 & 4.95    & 176.64 & 86.64 \\
\enddata
\tablecomments{$\mathrm{EVPA}_{\rm wrap} = \mathrm{EVPA} + 180\degr$ remaps the measured EVPA from the standard range $(-90\degr,+90\degr]$ into $(90\degr, 180\degr]$, exploiting the $180\degr$ periodicity of linear polarization. This convention avoids negative entries and is used solely for display purposes. The projected magnetic field position angle (B-field column) is then $\mathrm{PA}_{\rm B} = \mathrm{EVPA}_{\rm wrap} - 90\degr \equiv \mathrm{EVPA} + 90\degr$, measured from north to east.}
\end{deluxetable*}

An independent estimate of the magnetic field in the radio core can be obtained from the synchrotron self-absorption (SSA) turnover, assuming a homogeneous emitting region with an angular size approximately twice the K-band VLBA resolution and a Doppler factor $\delta \sim 1$, consistent with a mildly relativistic, nearly edge-on jet \citep{gallo2013}. Under these assumptions, the SSA magnetic field strength is given by \citep{Marscher1983}:
\begin{equation}
\label{eq:BSSA}
B_{\mathrm{SSA}} = 10^{-5} b(\alpha_{\mathrm{thin}})\, S_{\mathrm{p}}^{-2}\, \theta^4\, \nu_{\mathrm{p}}^5
\left( \frac{\delta}{1+z} \right)^{-1} \ \mathrm{G},
\end{equation}
where $S_{\rm p}$ is the peak flux density in Jy, $\theta$ is the angular diameter of the emitting region in mas, $\nu_{\rm p}$ is the peak frequency in GHz, $\delta$ is the Doppler factor, $z$ is the redshift, and $b(\alpha_{\rm thin})$ is a slowly varying numerical factor tabulated in \citet{Marscher1983}. For the optically thin spectral index $\alpha_{\rm thin} = -0.96 \pm 0.04$ obtained directly from our VLA-only peaked-spectrum fit~\citep[e.g.,][]{callingham2017}, interpolation of the \citet{Marscher1983} table yields $b(\alpha_{\rm thin}) \simeq 3.69$. The remaining parameters are taken directly from the peaked-spectrum fit and known source properties: $S_{\rm p} = 0.12$~Jy, $\nu_{\rm p} = 3.42$~GHz, $\delta = 1.0$, and $z = 0.017$ (corresponding to the source distance of 74.2~Mpc; annotated in Figure~\ref{fig:ssa_fit}). Adopting $\theta \sim 0.6$~mas, approximately twice the K-band VLBA restoring beam (Table~\ref{tab:Observation}), Equation~(\ref{eq:BSSA}) yields $B_{\rm SSA} \sim 171$~mG. We note that $\theta$ should strictly correspond to the size of the emitting region at the turnover frequency, which lies closest to our C-band (4.87~GHz) VLBA data. A Gaussian fit to the C-band core in the 2024 August epoch, contemporaneous with the VLA spectrum, returns a size consistent with the restoring beam and no significant deconvolved component, indicating that the core remains unresolved at this frequency. The emitting-region size at the turnover therefore cannot be measured directly, and since $B_{\rm SSA} \propto \theta^4$ it remains the dominant uncertainty in this estimate. We emphasize that this expression is derived for a homogeneous, self-absorbed sphere, for which $\alpha_{\rm thick} = 5/2$ by construction. Since our measured $\alpha_{\rm thick}$ departs from that value, $B_{\rm SSA}$ should be regarded as an order-of-magnitude estimate. Repeating the calculation with the parameters of the $\alpha_{\rm thick} = 5/2$ fit ($S_{\rm p} = 0.13$~Jy, $\nu_{\rm p} = 1.86$~GHz) gives $B_{\rm SSA} \sim 6$~mG, so the two limiting cases bracket a range of roughly $6$--$171$~mG. This range is consistent with the magnetic field strength estimated from the rotation measure analysis ($\langle B_\parallel \rangle \sim 1$--$100$~mG), considering uncertainties in the electron density, path length, and source size. The consistency between these independent estimates suggests that the Faraday rotating plasma is located close to the synchrotron-emitting jet, consistent with a magnetized sheath surrounding the inner parsec-scale jet. This sheath represents the inner, more compact component of the stratified Faraday-rotating medium; the outer, more turbulent component of this medium is responsible for the stronger depolarization observed at lower frequencies (Section~\ref{sec:vla_evpa}). If the source is inhomogeneous or multiple unresolved regions contribute to the emission, a filling factor $f<1$ would increase the inferred magnetic field strength by a factor $f^{-\eta}$, where $\eta \ge 0$. 

The EVPAs, after unwrapping, indicate a high-frequency magnetic field orientation (K through Q bands; B-field PA $\sim80$--$90^\circ$) that is offset from the NE--SW jet axis (PA $\sim40^\circ$) by $\sim40$--$50^\circ$. The stability of this orientation across the K, Ka, and Q bands ($\lesssim10^\circ$ variation; Table~\ref{tab:evpa_lambda2}) demonstrates that the field is ordered rather than tangled, while the offset from the jet axis is consistent with oblique shocks or shear near the jet base, as commonly observed in young or newly reactivated AGN jets (see Section~\ref{sec:vla_multiband_polarization} and references therein). Together these results support the interpretation that 1ES\,1927+654 is exhibiting the emergence of an ordered magnetic field during the early stages of jet formation.

\begin{figure*}[ht!]
    \centering
    \includegraphics[width=0.8\textwidth, clip, trim=1.5cm 4cm 1.5cm 3cm]{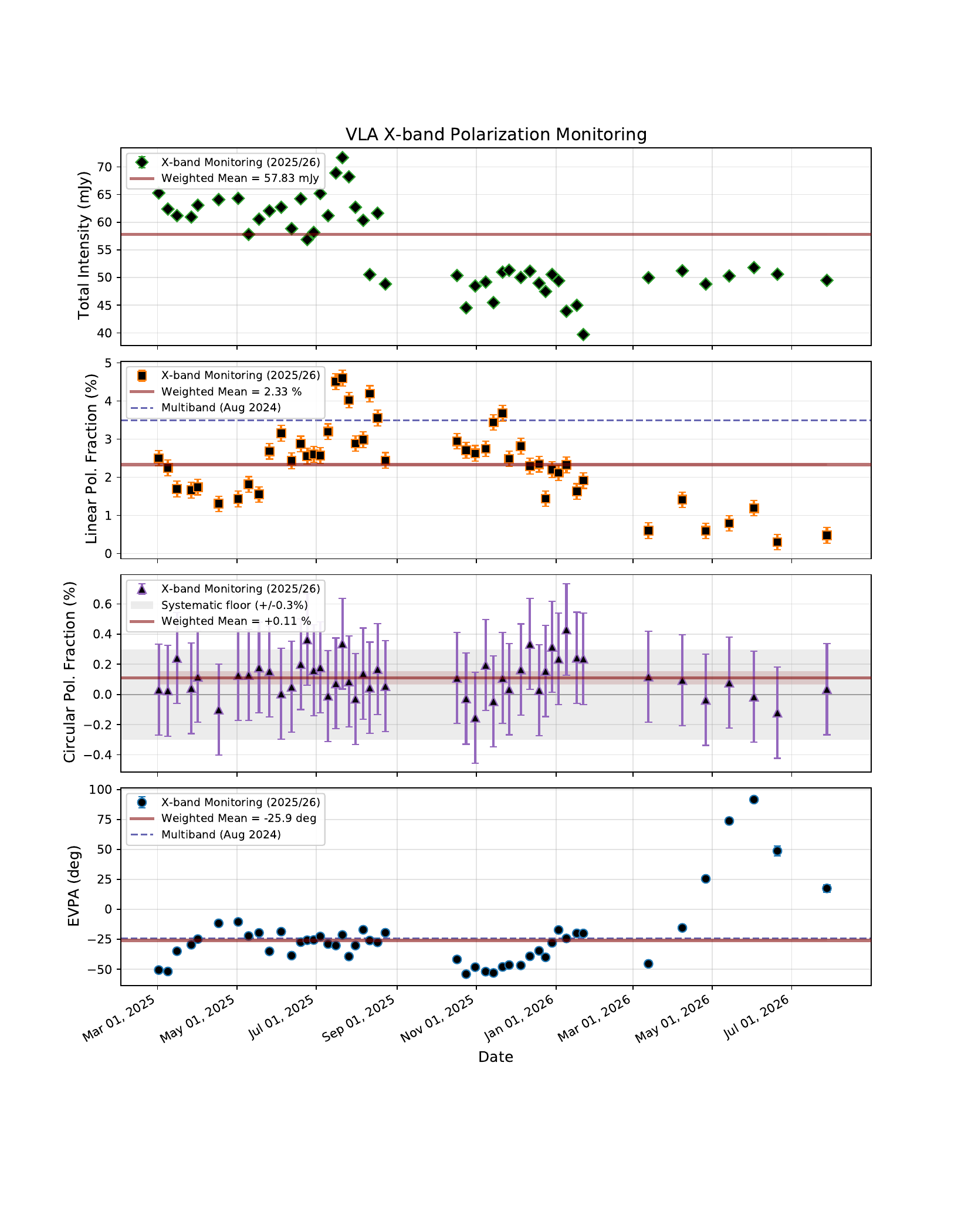}
    \caption{Temporal evolution of the X-band polarization properties of 1ES\,1927+654 from 2025 March to 2026 July. \textit{Top panel:} Total intensity (Stokes $I$, mJy) as a function of time, showing a general decline from $\sim$65--70~mJy in early 2025 to $\sim$45--55~mJy by late 2025, followed by a stable plateau near $\sim$50~mJy through the final months of the campaign. \textit{Second panel:} Linear fractional polarization ($m_{\rm pol}$) over the same period. Error bars correspond to $\sigma_m$, computed from the thermal noise in the Stokes $Q$ and $U$ images together with a systematic floor of $0.2\%$. \textit{Third panel:} Circular polarization, $V/I$. The shaded band marks the $\pm0.3\%$ systematic floor set by residual $R$/$L$ gain amplitude errors; all measurements lie within it, and circular polarization is not detected at any epoch. \textit{Bottom panel:} EVPA over the same period, unwrapped under the standard minimal-rotation convention. Values are therefore accumulated angles and may lie outside the $(-90^\circ, +90^\circ]$ range of the measured quantity, which is defined only modulo $180^\circ$. The dashed navy lines indicate the 2024 August multiband VLA reference values; the solid maroon lines show the weighted means. The 2026 epochs show a smooth EVPA rotation of $\sim137^\circ$ over 81 days followed by a partial return, at stable total intensity and suppressed fractional polarization, discussed in Section~\ref{sec:vla_xband_evpa}.}
    \label{fig:X_band_pol_evolution}
\end{figure*}

\begin{figure*}[ht!]
    \centering
    \includegraphics[width=1\textwidth]{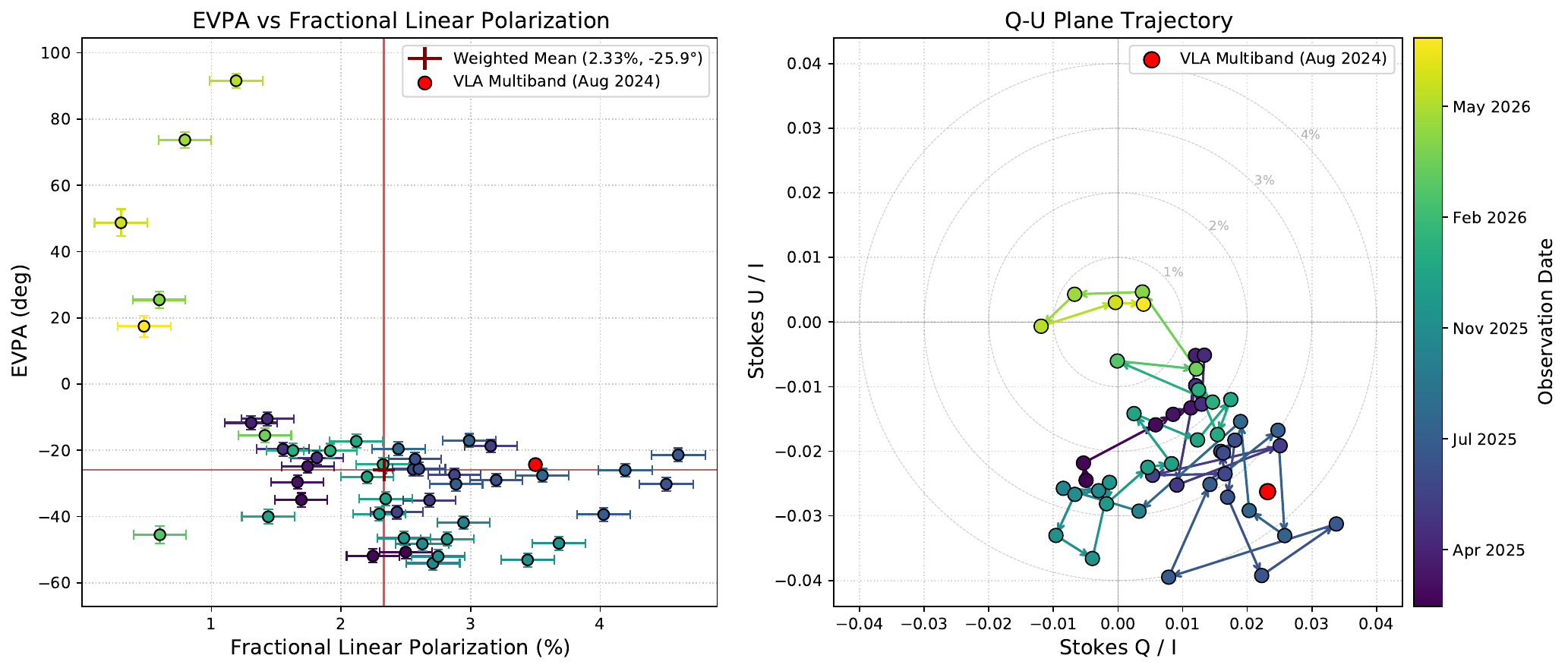}
    \caption{\textit{Left:} EVPA as a function of fractional linear polarization for VLA X-band monitoring of 1ES\,1927+654 from March 2025 through July 2026. Each point is color-coded by observation date (colorbar at right). Solid maroon lines indicate the weighted mean fractional polarization ($2.33\%$) and EVPA ($-25.9^\circ$). The 2026 epochs (light green through yellow) occupy a distinct region of the diagram, combining the lowest fractional polarizations in the campaign ($\sim0.30$--$1.4\%$) with the largest EVPA values, reaching $+91.6^\circ$ on June 2. The 2025 epochs instead scatter about the weighted mean. \textit{Right:} Stokes $Q/I$--$U/I$ plane trajectory for the same epochs. Dashed circles indicate constant fractional polarization levels of 1\%, 2\%, 3\%, and 4\%. The 2026 epochs trace a coherent counter-clockwise sweep followed by a reversal, at consistently low polarized fraction. The red point shows the August 2024 multiband reference.}
    
    \label{fig:EVPA_PolFrac}
\end{figure*}

\subsection{VLA X-band Monitoring}

\subsubsection{Fractional Polarization}
Figure~\ref{fig:X_band_pol_evolution} (second panel) shows the evolution of the fractional polarization of 1ES\,1927+654 over time at X band. The orange squares represent measurements from March 2025 to July 2026, with vertical error bars indicating the polarization uncertainty $\sigma_m$. Horizontal error bars are negligible since each epoch corresponds to a single observation day. For reference, the dashed navy line shows the fractional polarization from the August 2024 VLA multiband observations.

From March to May 2025, the fractional polarization gradually decreased from $\sim2.5\%$ to $\sim1.4\%$. Following this initial decline, the polarization fraction rose through mid-summer 2025, reaching a maximum of $\sim4.6\%$ in late July and early August, before declining again toward late August. Through late 2025 the fractional polarization remained in the range $\sim1.4$--$3.7\%$, and from March 2026 onward it dropped to persistently low values of $\sim0.3$--$1.4\%$, reaching a minimum of $\sim0.30\%$ on 2026 June 20, the lowest value in the campaign. Such variability in fractional polarization has been observed in other compact AGN jet cores, where changes on timescales of days to weeks are commonly attributed to evolving opacity, shock propagation, or changes in the magnetic field ordering within the innermost jet regions \citep{Zobnina2023, Hodge2018, marscher2008}. The sustained suppression observed through 2026 persists over considerably longer timescales, and we discuss it together with the contemporaneous EVPA rotation in Section~\ref{sec:vla_xband_evpa}.

To characterize the variability quantitatively, we estimated timescales $\Delta t$ for the fluctuations in fractional polarization. Change-point analysis identifies multiple short-term variations on timescales of a few days to weeks, while a discrete correlation function analysis yields a decorrelation timescale of $\sim29$~days. Similar decorrelation timescales have been reported in VLBI polarimetric monitoring of compact AGN jets \citep{Zobnina2023, Kravchenko2017, Hodge2018}.

From March through approximately October 2025, the total intensity and fractional polarization vary in a broadly similar fashion (Figure~\ref{fig:X_band_pol_evolution}, top and second panels), making it difficult to disentangle intrinsic polarization changes from variations driven by the evolving total flux density during this period. Such a correlation is expected when a single physical change drives both the emissivity and the degree of field ordering within the beam. Two cases are common. First, if the observed emission is the sum of a steady, weakly polarized core and a variable, more highly polarized component, then brightening of the latter raises both the total flux density and the net ordered fraction, so that $m_{\rm pol}$ and $I$ track one another. Second, shock compression simultaneously enhances the synchrotron emissivity and orders an initially tangled field, again producing a joint rise. From November 2025 onward, however, the total intensity settles onto a comparatively stable plateau near $\sim50$~mJy, while the fractional polarization and EVPA continue to vary substantially, including the extended low-polarization, large-EVPA-rotation epochs of 2026 discussed in Section~\ref{sec:vla_xband_evpa}. This decoupling between a stable total intensity and a strongly varying polarization signal during the final months of the campaign suggests that the polarimetric variability in this period is not primarily driven by changes in the overall flux density. Conversely, when the total intensity is stable while $m_{\rm pol}$ and the EVPA vary substantially, the variability cannot be attributed to changes in emissivity and must instead reflect changes in the magnetic field geometry or in the Faraday-active medium. In addition, the gradual decline in total intensity observed through mid-to-late 2025, from $\sim$65--70~mJy to $\sim$50~mJy, is suggestive of the fading of extended, more highly polarized jet components as they expand and cool adiabatically, with the beam becoming increasingly dominated by the less-polarized compact core.

\subsubsection{Circular Polarization}\label{sec:circular_pol}

The third panel of Figure~\ref{fig:X_band_pol_evolution} shows the circular polarization, $m_{\rm circ} = V/I$, measured at the same pixel as the total and linearly polarized intensities. No epoch shows a significant detection: the individual measurements scatter about a weighted mean of $+0.11\%$ and lie within the $\pm0.3\%$ systematic floor set by residual $R$/$L$ gain amplitude errors, which we adopt as a conservative upper limit. The absence of detectable circular polarization is consistent with the low levels ($\lesssim0.5\%$) typically found in compact AGN cores at centimetre wavelengths, and argues against a strong contribution from Faraday conversion in the emitting region. Since the leakage terms that would produce spurious $V$ also corrupt $Q$ and $U$ at a comparable level, the non-detection additionally supports the reliability of the linear polarization measurements reported above.

\subsubsection{EVPA Evolution}\label{sec:vla_xband_evpa}

In the bottom panel of Figure~\ref{fig:X_band_pol_evolution}, we present the EVPA as a function of time for the same epochs as the fractional polarization measurements. The EVPA at each epoch was computed from the Stokes $Q$ and $U$ values at the peak-intensity pixel. Because the EVPA is defined only modulo $180^\circ$, the measured values were unwrapped by applying the standard minimal-rotation convention to the full time-ordered series, adding integer multiples of $180^\circ$ so that the change between consecutive epochs is less than $90^\circ$ \citep{Hovatta2012, Kiehlmann2016}. Prior to the 2026 rotation the EVPA varies stochastically about its mean with no significant secular drift: a weighted linear fit to the 41 epochs before 2026 March gives $-0.025 \pm 0.019$~deg~d$^{-1}$, with an rms scatter of $11.9^\circ$ about that fit.

To identify coherent rotations objectively we applied the criteria of \citet{Kiehlmann2016}, requiring consecutive EVPA changes exceeding $5^\circ$ and $2\sigma$, with a consistent sense of rotation and no gap longer than 45 days. Four episodes are recovered across the campaign. Two are short and modest: $+22.3 \pm 2.9^\circ$ over 11 days in 2025 July--August, and $+22.8 \pm 3.0^\circ$ over 10 days spanning 2025 December to 2026 January. The remaining two constitute the 2026 event. The EVPA begins at $-45.5^\circ$ on March 13 and rises monotonically through $-15.5^\circ$ (April 8), $+25.4^\circ$ (April 26) and $+73.7^\circ$ (May 14), reaching $+91.6^\circ$ on June 2: a rotation of $+137.1 \pm 3.3^\circ$ over 81 days, significant at $41\sigma$ and comprising four consecutive same-sense intervals. The rate accelerates from $\sim1.2^\circ$~d$^{-1}$ to $\sim2.7^\circ$~d$^{-1}$ over the rise before decelerating to $\sim0.9^\circ$~d$^{-1}$ in the final interval. The EVPA then returns, reaching $+48.7^\circ$ on June 20 and $+17.5^\circ$ on July 28, a change of $-74.1 \pm 3.9^\circ$ over 56 days at $19\sigma$. Every consecutive step is well below $90^\circ$, so the sequence is continuous and requires no discontinuity at any epoch. The 2026 rotation exceeds the largest 2025 episode by a factor of six in amplitude and by nearly an order of magnitude in duration. Throughout this rotation the total intensity remains stable at $49$--$52$~mJy (March 13: 50.0~mJy; June 2: 51.8~mJy; July 28: 49.5~mJy), while the fractional polarization remains suppressed $\lesssim1.4\%$, reaching a minimum of $\sim0.30\%$ on June 20. This behavior is visible in the Q--U plane trajectory (Figure~\ref{fig:EVPA_PolFrac}, right panel) as a coherent sweep followed by a reversal, traced at consistently low polarized fraction. EVPA rotations of comparable magnitude and duration have been reported in monitoring campaigns of compact AGN jets and blazars, where they are variously attributed to geometric effects, propagating shocks, or changes in the Faraday screen \citep{marscher2008, Agudo2012, Kravchenko2017, Cohen2014}.

At least two interpretations are possible for this rotation. First, it may reflect a transient opacity change near the SSA turnover (Figure~\ref{fig:ssa_fit}), which lies at $\nu_{\rm peak} \approx 3.4$~GHz, only a factor of $\sim3$ below the X-band observing frequency of 10~GHz. Near $\tau \sim 1$, both the EVPA and the degree of polarization are sensitive to the relative contributions of optically thick and thin emission regions; a transient shift in opacity could simultaneously rotate the EVPA and suppress $m_{\rm pol}$, as observed. However, an opacity-driven change of sufficient magnitude to produce a $\sim137^\circ$ EVPA rotation would generally be expected to also modify the total flux density \citep{pacholczyk1970}, which remains stable at $\sim49$--$52$~mJy throughout, arguing against this scenario as the primary driver. Second, and more consistent with the total intensity stability, the sustained rotation may reflect a genuine reorganization of the magnetic field geometry within the compact emitting region, driven by a disturbance propagating through the core. This may be a newly ejected component or an internal shock. As the disturbance traverses the core, superposition of its magnetic field with the pre-existing ordered field rotates the net EVPA and dilutes the fractional polarization (as fields of differing orientations partially cancel); the smooth, progressive character of the observed rotation is consistent with a crossing over several months rather than instantaneously, and the subsequent partial return of the EVPA corresponds to the disturbance leaving the emitting region, indicating a transient rather than secular change \citep{marscher2008, Cohen2018, Beuchert2018, Marscher2014, Kravchenko2017, Pushkarev2023}.

Simultaneous multi-frequency VLA polarimetric observations, which have been approved, will repeat the August 2024 multi-band measurement and allow a direct comparison of the frequency-dependent polarization before and after the rotation. They will be presented in a future study.

\section{Conclusions}\label{sec:conclusions}

\subsection{VLBA Imaging and Jet Structure}

We obtained new multi-frequency VLBA observations in L, C, X, and K bands, providing milliarcsecond-scale resolution of the nuclear region of 1ES\,1927+654. K-band images continue to reveal the evolving bipolar two-component structure first reported in \citet{meyer2025}. The higher-resolution K-band imaging additionally reveals a bridge of residual emission connecting the two compact peaks, suggesting a continuous outflow structure rather than discrete independently ejected knots. From August 2024 onward, X-band imaging confirms the same bipolar morphology at larger angular separation, with position angles consistently falling within $\sim35^\circ$--$45^\circ$ across all epochs and both frequency bands, indicating outward propagation of the same jet system along a stable axis. The spectral behavior of the VLBA core across L--K bands is consistent with a GHz-peaked spectrum, characteristic of a young, compact, self-absorbed synchrotron source.

\subsection{VLA Multi-band Flux Density, Spectral, and Polarization Properties}

VLA flux densities closely match contemporaneous VLBA measurements above 8~GHz, demonstrating that the radio emission is dominated by a compact core. At lower frequencies, a modest VLA flux excess over VLBA indicates some contribution from intermediate-scale emission resolved out on VLBA baselines, but no significant large-scale extended emission is detected, consistent with a very young jet that has not yet propagated to large spatial scales. VLITE+VLA multi-band observations reveal a spectral turnover at $\sim3$~GHz, interpreted as synchrotron self-absorption in an inhomogeneous source, confirming the GPS classification of this newly formed compact jet. Linear polarization is detected above $\sim3$~GHz, with fractional polarization rising from $\sim0.2\%$ to $\sim6\%$ across S to Q bands. The EVPA exhibits non-linear $\lambda^2$ behavior requiring at least two Faraday rotation components, revealing a stratified magnetized environment: a high-frequency RM ($\mathrm{RM}_{\rm X-Q} = -452$~rad~m$^{-2}$) tracing an inner magnetized sheath, and a low-frequency RM ($\mathrm{RM}_{\rm S-X} = -108$~rad~m$^{-2}$) reflecting a more turbulent outer Faraday screen. Parsec-scale magnetic field strengths of $\langle B_\parallel \rangle \sim 1$--$100$~mG are inferred from the RM analysis, broadly consistent with an independent SSA-based estimate of $B_{\rm SSA} \sim 6$--$171$~mG.

\subsection{VLA X-band Monitoring and Polarimetric Evolution}

Long-term VLA X-band monitoring captures the evolution of fractional polarization and EVPA on weekly-to-monthly timescales and exhibits variability in both. The fractional polarization decorrelates on a timescale of $\sim29$~days. Prior to 2026 the EVPA shows no significant secular drift, varying stochastically about its mean with an rms scatter of $11.9^\circ$. The most dramatic polarimetric event in the monitoring campaign is a smooth EVPA rotation of $\sim137^\circ$ over $\sim81$~days in 2026, followed by a partial return of $\sim74^\circ$, while the total intensity remains stable to within a few per cent and the fractional polarization stays suppressed $\lesssim1.4\%$. The stability of the total flux density argues against a simple opacity change and is consistent with a propagating disturbance traversing the compact emitting region \citep{marscher2008, Cohen2018, Beuchert2018}. Circular polarization is not detected at any epoch, with $|V|/I$ consistent with zero at the $\sim0.3\%$ level. Simultaneous multi-frequency polarimetric follow-up observations, which have been approved, will show whether the field configuration was permanently altered.

The combined VLBA and VLA observations show that 1ES\,1927+654 has transitioned from an X-ray--dominated, non-jetted state to a radio-loud, jet-producing AGN, with the appearance of a GHz-peaked radio spectrum, the first detection of linear polarization, and the emergence of compact VLBA jet components all occurring alongside the rise in soft X-ray flux and the emergence of Fe K emission~\citep{Sadaula_2026arXiv260705246S}. The frequency-dependent rise in fractional polarization, the two distinct Faraday rotation regimes, and the stable high-frequency EVPA together reveal a layered jet structure consisting of an inner ordered jet surrounded by a turbulent magnetized sheath, consistent with magnetically driven jet models that predict large-scale polarization properties should retain memory of the inner jet magnetic geometry \citep{Blandford1979, Blandford2019}. The large EVPA rotation observed in our monitoring campaign further establishes 1ES\,1927+654 as a unique laboratory for studying the early polarimetric evolution of a newly formed relativistic jet in real time, and future simultaneous multi-frequency polarimetric monitoring will allow us to fully characterize the magnetic field structure of this layered system.

\begin{acknowledgments}

This work was supported by the National Science Foundation under Grant No. AST-2407801 and by the United States–Israel Binational Science Foundation (BSF) under Grant No. 2023752. S.L. acknowledges support from NASA under award number 80GSFC21M0002. N.S. acknowledges support from the Simons Foundation (grant MP-SCMPS-00001470). Basic research in Radio Astronomy at the U.S.\ Naval Research Laboratory is funded by 6.1 Base funding. Construction and installation of the VLA Low-band Ionosphere and Transient Experiment (VLITE) was supported by the NRL Sustainment Restoration and Maintenance fund. The National Radio Astronomy Observatory is a facility of the National Science Foundation operated under cooperative agreement by Associated Universities, Inc. We acknowledge Phil Cigan (USNO) and the MultiColorFits package, which was used to create the VLBA image figures \citep{Cigan2019}. We also thank Andrew J. Sargent (USNO) for his support in building the VLA polarization calibration pipeline. 
 
\end{acknowledgments}

%
\vspace{5mm}
\facilities{Very Large Array (VLA); VLA Low-band 
Ionosphere and Transient Experiment (VLITE); Very Long Baseline Array (VLBA); European VLBI Network (EVN); Arcminute Microkelvin Imager (AMI); enhanced Multi Element Remotely Linked Interferometer Network (e-MERLIN); Submillimeter Array (SMA); Swift X-ray Observatory}




\bibliography{1ES1927_polarization, 1ES1927_Radio_Outburst}{}
\bibliographystyle{aasjournal}



\end{document}